\documentclass[journal]{IEEEtran}
\usepackage{graphicx}
\usepackage{amsmath,amssymb}
\usepackage{subfigure}
\usepackage[font=scriptsize]{caption}
\usepackage{citesort}
\usepackage{fancyhdr}
\usepackage{mdwmath}
\usepackage{mdwtab}
\usepackage{balance}
\usepackage{xcolor}
\usepackage{bm}
\usepackage{amsthm}
\usepackage{multirow}
\usepackage{flafter}
\usepackage{caption}
\usepackage{setspace}
\usepackage{cite}
\usepackage{algpseudocode}
\usepackage{algorithm}
\usepackage{multirow}
\usepackage{flafter}
\DeclareMathOperator*{\argmax}{argmax}

\newtheorem{remark}{Remark}

\newtheorem{theorem}{Theorem}

\newtheorem{lemma}{Lemma}

\newtheorem{corollary}{Corollary}

\newtheorem{assumption}{Assumption}

\theoremstyle{definition}

\begin{document}
\title{Transmit Power Pool Design for Grant-Free NOMA-IoT Networks via Deep Reinforcement Learning}

\author{Muhammad~Fayaz,~\IEEEmembership{Student Member,~IEEE,} 		  Wenqiang~Yi,~\IEEEmembership{Member,~IEEE,} Yuanwei~Liu,~\IEEEmembership{Senior Member,~IEEE,} and Arumugam~Nallanathan,~\IEEEmembership{Fellow,~IEEE,}
\thanks{M. Fayaz, W. Yi, Y. Liu, and A. Nallanathan are with Queen Mary University of London, London, UK (email:\{m.fayaz, w.yi, yuanwei.liu, a.nallanathan\}@qmul.ac.uk).\newline
\indent M. Fayaz is also with the Department of Computer Science and  IT, University of Malakand, Pakistan.\newline
\indent Part of this work was submitted in IEEE International Conference on Communications (ICC), June, Canada, 2021 \cite{fayaz}.}
}

\maketitle

\begin{abstract}
Grant-free non-orthogonal multiple access (GF-NOMA) is a potential multiple access framework for short-packet internet-of-things (IoT) networks to enhance connectivity. However, the resource allocation problem in GF-NOMA is challenging due to the absence of closed-loop power control. We design a prototype of transmit power pool (PP) to provide open-loop power control. IoT users acquire their transmit power in advance from this prototype PP solely according to their communication distances. Firstly, a multi-agent deep Q-network (DQN) aided GF-NOMA algorithm is proposed to determine the optimal transmit power levels for the prototype PP. More specifically, each IoT user acts as an agent and learns a policy by interacting with the wireless environment that guides them to select optimal actions. Secondly, to prevent the Q-learning model overestimation problem, double DQN (DDQN) based GF-NOMA algorithm is proposed. Numerical results confirm that the DDQN based algorithm finds out the optimal transmit power levels that form the PP. Comparing with the conventional online learning approach, the proposed algorithm with the prototype PP converges faster under changing environments due to limiting the action space based on previous learning. The considered GF-NOMA system outperforms the networks with fixed transmission power, namely all the users have the same transmit power and the traditional GF with orthogonal multiple access techniques, in terms of throughput.
\end{abstract}

\begin{IEEEkeywords}
Double Q learning, Grant-free NOMA, Internet of things, Multi-agent deep reinforcement learning, Resource allocation
\end{IEEEkeywords}
\section{Introduction}
One of the main challenges to the next generation cellular networks is the provision of massive connectivity to explosively increased Internet-of-things (IoT) users, especially for uplink transmission. In current cellular networks, enabling multiple access with limited resources is an inherent problem. Fortunately, non-orthogonal multiple access (NOMA) with a new degree of freedom, namely the power domain, has been established as a promising technique for the solution of this problem \cite{wen}.
Some latest work investigating NOMA from different aspects can be found in \cite{zhao2020security} \cite{park2021sum} \cite{yang2021reconfigurable}.
Although grant-based (GB) has been widely studied, it fails to provide sufficient access to IoT users with short packets, since multiple handshakes are required before the transmission. Therefore, grant-free (GF) NOMA is proposed to enhance this connectivity.

In GF-NOMA, multiple IoT users transmit data in an arrive-and-go manner to the base station (BS) on the same time-frequency resource block (RB) without waiting for the BS to schedule and grant \cite{zhang2020deep}. However, the resource allocation problem in GF-NOMA is challenging as BSs commonly have no/partial information about the active users and their channel state information (CSI). Additionally, keeping enough power difference for successful successive interference cancellation (SIC) processes at the NOMA-enabled BS side is a tough practical challenge. As NOMA is heavily based on the received power difference among users \cite{8641425}, the effectiveness of such a solution is limited for GF schemes in the absence of a closed-loop power control \cite{9097306}. It is worth noting that user clustering and power allocation in NOMA is mainly depended upon their channel gain, which can be calculated via IoT users geographical information and practical statistic models. Such information can be acquired without information exchanges, which enables an open loop. Therefore, a prototype of transmit power pool for GF-NOMA in IoT can be designed, based on geographical information and statistic models, to ensure the received power level difference.
\subsection{Related Works}
To provide massive connectivity for IoT devices, the power domain NOMA is a practical solution. However, as compared to OMA, NOMA introduces some complications in resource allocation design from two aspects: 1) user clustering/grouping; and 2) power allocation. Therefore, systematic user clustering and an efficient power allocation algorithms are required to ensure SIC processes at the receiver. Furthermore, within a cluster, each user needs to decode other users' information which increases the complexity and energy consumption at the receiver. Moreover, in uplink transmission, if an error occurs during the SIC process for a single user, decoding fails for all other users. Therefore, a significant channel gain difference is preferable, otherwise, the desired functionalities of the power domain concept cannot be achieved. Besides, each user in the network needs to report its channel gain back to BS, so NOMA is sensitive to capturing such measurements. Despite the complicated resource allocation design, NOMA still has tremendous advantages over OMA, especially in terms of connectivity and throughput. Next, we present a brief overview of existing works investigating NOMA with GF transmission.
Cellular IoT networks commonly use two types of random-access protocols known as GB and GF access protocols. In GB transmission, users or devices process a four-step random access protocol before the data transmission \cite{zhang2020deep}, \cite{9097306}.  GB-NOMA access leads to high signaling overhead and long latency, which makes GF-NOMA inevitable. In GF schemes, whenever users wants to transmit their data, neither explicit grant nor schedule request is needed that significantly reduces latency and signaling overhead. GF schemes are well suited for one typical IoT use case, named massive machine-type communications (mMTC)\cite{ZTE}, \cite{Lenovo}. However, if two or more users select the same resource for transmission, a collision occurs. Under this scenario, the receiver is unable to decode the data of users sharing the same RB.

Some GF schemes based on conventional optimization approaches are discussed in  \cite{7972955}, \cite{abbas2018novel}. Authors in \cite{7972955}, \cite{abbas2018novel} have split the cell area into partitions while dividing the users and sub-channels into the same number of partitions. To prevent collisions among MTC users, they used orthogonal resource in different layers.

Applying partial network observations and uniform resource access probabilities expropriate the conventional optimization approaches for GF transmission, especially for long-term communications with time-varying channels. Deep reinforcement learning (DRL) is applied to improve the GF transmission and to obtain better resource allocation with near-optimal resource access probability distribution \cite{zhang2020deep}. DRL based resource allocation for GF transmission is given in \cite{zhang2020deep}, \cite{chang2018distributive}, \cite{xu2017dynamic}, \cite{wang2018deep}. To reduce collisions, authors in \cite{zhang2020deep} designed users and sub-channel clusters in a region, where the number of users compete in a GF manner for several available sub-channels in each region. The formulated long-term cluster throughput problem is solved via DRL based GF-NOMA algorithm for optimal sub-channel and power allocation. As compared to slotted ALOHA NOMA, the DRL based GF-NOMA algorithm shows performance gain in the system throughput. Similarly, a recent work \cite{chang2018distributive} investigates the problem of channel selection of secondary user, performing channel selection from sensing the history of the secondary user through DRL.  In \cite{xu2017dynamic}, authors modelled users as cluster heads to maximize capacity and to ensure delay requirements via multi-agent learning algorithm. Data rate maximization and a number of long-term successful transmissions problem are investigated in \cite{wang2018deep} using Q-learning algorithm.
\begin{figure*}[t!]
	\centering
	\includegraphics[width = 1 \linewidth,keepaspectratio]{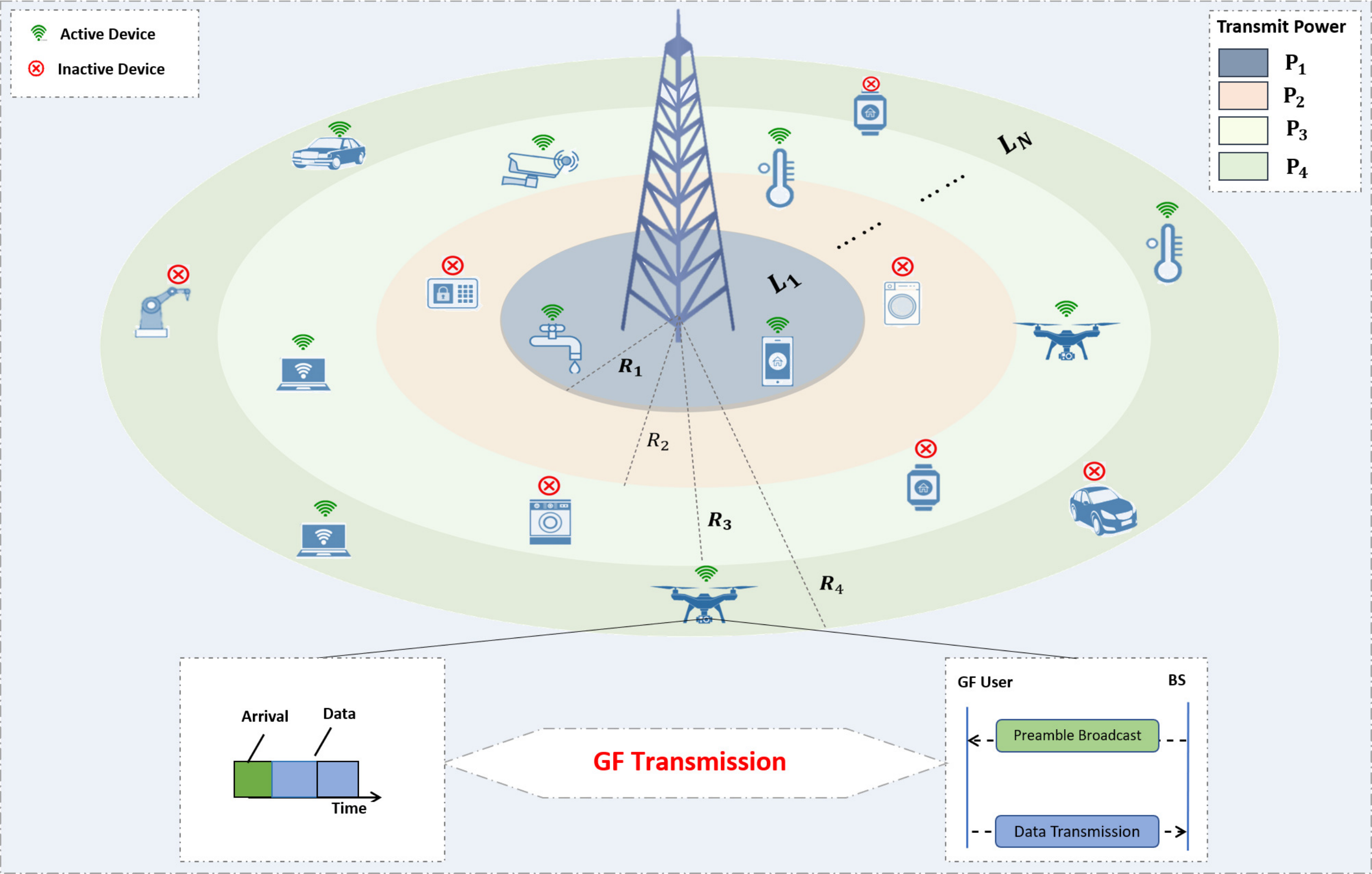}\\
	\caption{Illustration of Grant-Free NOMA IoT Networks: The top figure represents a subset of IoT devices $N_t$ active in a time slot $t$ that transmit in GF manner. The entire cell is divided into $L$ concentric layers with the same radius difference, IoT users in different layers are able to acquire their aiming transmit power ($P1, P2, \cdots$) from the transmit power pool in advance for helping GF-NOMA transmission. The bottom sub-figures shows GF procedure, in which the BS broadcasts a preamble including a well-designed transmit power pool, and IoT user transmit data without any prior handshake.}\label{fig:image2}
\end{figure*}
\subsection{Motivation and Contributions}
The conventional GF-NOMA is not suitable for IoT networks because users transmit at fixed power and to find the optimal transmit power for each user, it needs a closed-loop power control. Thus, GF-NOMA with fixed power control introduces additional signal overhead and leads to energy consumption. To enable GF transmission with open-loop power control and less signalling overhead, a prototype of transmit power pool can be designed based on geographical region. This prototype power pool can enable IoT users to transmit with low power consumption and reduces computational complexity by preventing closed-loop power control.
The aforementioned research contributions considered solutions for mitigating the problem of collisions and enhancing GF transmission by both conventional and machine learning methods. However, in these approaches, BSs need to collect information about users that include instant users rates, the number of active users in the network and the best grouping policy that BS broadcast to all users. Such prerequisites increase complexity at the BS side due to massive information exchange between BSs and IoT users \cite{9205230}.
In resource optimization problems, ML algorithms have several advantages over conventional optimization approaches. The conventional optimization approaches lead to complexity and high cost as the number of parameters to be configured increases. Traditional optimization algorithms are often prone to parameter selection, and heuristics must be run from scratch every time there is a small change in the system model, such as when new users are added. In other words, a small change in any system parameter requires the entire algorithm to be run from scratch each time \cite{9154358}. Moreover, with conventional methods, the open-loop power control is difficult to be achieved, and the received power level difference cannot be guaranteed. It is noteworthy that resource allocation in wireless networks is an non-deterministic polynomial-time hard (NP-hard) problem \cite{9167258}. In addition, calculating optimal solution is a combinatorial optimization problem, which is mathematically intractable with increasing network size due to partial state observations. However, machine learning (ML) can be used to solve NP-hard optimization problems more efficient as compared to traditional optimization approaches \cite{9097306}. ML methods observe the patterns in data as a substitute for relying on equations and models for near-optimal and best possible decisions. The ML-based algorithms are desirable in 5G and beyond wireless communications, especially for mMTC, as the complexity of such processes increase exponentially with the number of users \cite{ahsan2020resource}. Besides, reinforcement learning (RL) has the potential of taking decisions and perform learning simultaneously, which is one of the ideal characteristics for the applications of wireless communication \cite{luong2019applications}. Therefore, we adopt an ML-based algorithm due to its potential to offer excellent approximate solutions while interacting with a huge state and action spaces.
Furthermore, due to the unavailability of realistic datasets, RL algorithms are able to generate datasets during simulation (online) to learn hidden patterns for optimal decisions.

Based on the aforementioned issues, we propose a ML-based scheme to address the issue of complexity at BSs by creating a power pool associated with users location information which is not yet considered in the literature. In this paper, we propose a multi-agent deep Q network (DQN) and double DQN (DDQN) based GF-NOMA algorithm for prototype power pool design, where the BS broadcasts this pool to all IoT users so as to avoid acquiring CSI. Each IoT user can randomly select one power level for transmission that reduces complexity at BS and avoid massive information exchange between IoT user and the BS.
The power selection from this well-designed prototype power pool guarantees distinct received power levels at the BS for successful SIC processes and reduces collision probabilities by allowing pilot sequence reusing.
To the best of the author's knowledge, this is the first work to design a power pool for GF-NOMA via multi-agent reinforcement learning (MARL). In a nutshell, this work provides the following four major contributions.
\subsubsection{Novel Power Pool Framework for GF-NOMA}
We consider uplink transmission in IoT networks with the traffic model of packets following the Poisson distribution. Further, we divide the cell area into different layers and design a layer-based transmit power pool prototype via MARL. In the proposed framework, data transmitting IoT users select a transmit power based on their communication distance (layer) from the well-designed prototype power pool for GF-NOMA transmission, without any information exchange between IoT user and the BS. Based on the proposed framework, we formulate power and sub-channel selection for throughput optimization in GF-NOMA systems, an optimization problem.
\subsubsection{Novel Designs for the MARL}
We implement a multi-agent DQN aided GF-NOMA algorithm to acquire the optimal transmit power levels for each layer. In the proposed multi-agent DQN model, the IoT user acts as a learning agent and interacts with the environment. After learning from its mistakes, finally, the IoT users in each layer find out the optimal transmit power level that maximizes network throughput. We adopt a multi-agent DDQN based GF-NOMA algorithm to prevent the action values overestimation problem encountered by the conventional Q-learning model.
\subsubsection{Long-Term Resource Allocation}
With the aid of MARL methods, we find the
optimal resource allocation (transmit power and sub-channel) strategy, where multi-agent DDQN based GF-NOMA algorithm with learning rate $\alpha =0.001$ provides better system throughput and finds the optimal transmit power levels (prototype power pool) for each layer. We show the advantages of multi-agent DDQN over traditional multi-agent DQN for GF-NOMA IoT networks. In particular, we demonstrate that, compared to the multi-agent DQN, multi-agent DDQN converges to a more stable and optimal solution (optimal resource allocation). Moreover, we showed that the algorithm with the prototype power pool converges with fewer training episodes as compared to the algorithm with available power levels.

\subsubsection{Performance Gain of GF-NOMA over GF-OMA}
Simulation results verify that multi-agent DDQN based GF-NOMA outperforms conventional GF-OMA based IoT networks with 55\% performance gain on system throughput. Additionally, transmit power allocated to IoT users from the available power pool achieves 37.7\% more throughput as compared to fixed power allocation strategy.

The rest of the paper is organized as follows. The system model is presented in Section II. The multi-agent DRL-based GF-NOMA user's power level and sub-channel selection algorithms are given in Section III. Numerical results and discussion are shown in Section IV. Conclusions are drawn in Section V.
\begin{table*}[htb!]
	\footnotesize
	\centering
	\caption{Table of notations}
	\label{tab11}
	\begin{tabular}{l|l|l|l}
		\hline
		\hline
		\textbf{Symbol} &  \textbf{Definition} &   \textbf{Symbol} &  \textbf{Definition} \\ \hline
		\text{$\mathbf{U}$} &  The set of entire IoT users &\text{$N_t$} &  Number of active IoT users in a time slot $t$   \\ \hline
		\text{$R$}  &   {Radius of the cell}& \text{$L$} & {Number of layers in cell area}  \\ \hline	
		\text{$\boldsymbol{V_l}$} & {Set of IoT users in layer $l$}& \text{$M$ } & {Number of sub-channels}  \\ \hline
		\text{$B_s$ } & {Sub-channel bandwidth}&   {$P_l$} & {Transmit power level for set of IoT users in layer $l$}  \\ \hline
		\text	{$r_{i,j}$} & {Communication distance between IoT user $j$ and the BS}&    \text{$n_0$} & {Additive white Gaussian noise }\\ \hline
		\text{$P^r_{i,j}(t)$} & {Received power of user $j$ via sub-channel $i$} &   {$\gamma_{i,j}(t)$} & {SINR of users $j$ on sub-channel $i$}\\ \hline
		\text{$\mathbf{P}_{pool}$} & {Prototype of transmit power pool}&
		{$E_p$} & {Number of transmit power levels in the designed power pool}\\\hline
		\text{$P_{BS}$} & {Transmit power of BS}&
		{$R_{th}$} & {Date rate threshold requirement for successful SIC}\\\hline
		\text{$k_{i,j}(t)$} & {Sub-channel selection variable for user $j$ and sub-channel $i$}&
		{$R_{i,j}$} & {Data rate of user $j$ on sub-channel $i$}\\\hline
		\text{$\mathbf{P}_t$} & {Matrix for transmit power levels}&
		{$N_p$} & {Number of available transmit power levels}\\\hline
		\text{$\mathbf{K}_t$} & {Matrix for sub-channel selection}&
		{$G$} & {Set of agents in MARL method}\\\hline
		\hline
	\end{tabular}
\end{table*}
\vspace{-0.2 cm}
\section{System Model}
We consider uplink transmission in IoT networks as shown in Fig.~\ref{fig:image2}, where a single BS is located at the origin of a circle with a radius $R$ and a set of IoT users $\mathbf{U}=\{u_1,u_2,...\}$ uploads messages to the central BS under GF-NOMA principles. More specifically, each IoT user independently selects its transmit power and NOMA cluster to send arrived data packets without waiting for any acknowledgement from the BS.
The entire cell is divided into $L$ concentric layers with different aiming received powers at the BS.
IoT users in different layers are able to acquire their aiming transmit power in advance for helping NOMA transmission. However, users in the same layer utilize the same transmit power for GF transmission. Hence, the prior information about IoT user's activity is not required that reduces the computational complexity and information exchange at the BS side.
An IoT user $j$ present in layer $l$, if its communication distance to the BS is $r_j$ and belongs to the set of users $\boldsymbol{V_l}\in \mathbf{U}= \{j|D_{l-1} < r_j \leq D_l\}$, where $D_0=0$ and $D_l= (R/L)l$. $D_0$ and $D_l$ define the boundaries of each layer. More specifically, $D_l= (R/L)l$ represents the upper boundary of each layer.
The aiming transmit power for the users in $\boldsymbol{V_l}$ is denoted by $P_l$. Moreover, the system bandwidth $B$ is equally divided into $M$ orthogonal sub-channels and thus each sub-channel has $B_s= \frac{B}{M}$ available bandwidth. Table \ref{tab11} summarizes the notations used in this research work.

\subsection{Traffic Model}
In GF NOMA principles, users transmit information in an arrive-and-go manner~\cite{zhang2017poc}. When a data packet arrives at one user, it sends this packet at the next time slot directly. Therefore, we assume the traffic model of packets in each time slot follows a Poisson distribution with an average arrival rate $\lambda$ (we used the Poisson distribution as a probability distribution to analyse the probability of the number of active users in a time slot, under this case, the probability of each user being active is the same). At one time slot $t$, the probability of the number of active users $n_t$ equalling to $N_t\ge 0 $ is given by
\begin{align}
\Pr\{n_t=N_t\} = \frac{\lambda^{N_t}\exp(-N_t)}{N_t!}.
\end{align}
This paper considers a general case that all users have the same priority.
\subsection{Path Loss Model}
We utilize a typical path loss model~\cite{8856258} with an intercept $C_I$ and path loss exponent $\alpha$. When the communication distance between one user and the central BS is $r$, the path loss law can be expressed as follows
\begin{align}\label{path_loss}
\mathcal{P}_L (r) = C_I r^{-\alpha}.
\end{align}

For an arbitrary active user, the probability density function (PDF) of the distance $r$ between it and the central BS is given by
\begin{align}
f_r (r) = \frac{2r}{R^2}.
\end{align}
Therefore, the path loss expression in \eqref{path_loss} obeys that $\Pr\{\mathcal{P}_L(r) = \mathcal{P}_L(x)\} = f_r(x)$.
\subsection{NOMA Transmission}
By using power-domain NOMA, BSs are able to serve multiple users with different receive power levels in the same sub-channel \cite{9139273}. For the central BS, the received information at the time slot $t$ is
\begin{align}
y (t) = \sum_{i=1}^{M} \sum_{j=1}^{N_{t,i}}\sqrt{P_{i,j}(t)\mathcal{P}_L(r_{i,j}(t))} h_{i,j}(t) x_{i,j}(t) + n_{0}(t),
\end{align}
where $N_{t,i}$ is the number of active users in the $i$-th sub-channel and $\sum_{i=1}^{M}N_{t,i}=N_t$.
The $\sum_{j=1}^{N_i}P_{i,j}(t)  \leq P_{max}$, $r_{i,j}$, $h_{i,j}$, and $x_{i,j}$ are the transmit power with imposed constraint for each user, communication distance, small-scale Rayleigh fading, and transmitted data for the $j$-th user~$(j \le N_{t,i})$ in the $i$-th sub-channel, respectively. The $n_{0}$ is the additive white Gaussian noise (AWGN). We consider a static channel at each time slot,
but the values of the aforementioned parameters are different for different time slots, i.e., these values for each user are constant during one time slot and change to new independent values for the next time slot. Obviously, different users can have different values in each time slot. Based on NOMA principles, the BS first decodes the user with strongest received power and then subtract its data from the received information via successive interference cancellation (SIC) techniques (we assume perfect SIC in this paper). After that, the BS turns to decode the user with the second strongest received power~\cite{8114722}. Therefore, the decoding order at the BS is in sequence to the received power.
However, considering heterogeneous QoS requirements could change this decoding order, but this consideration is beyond the scope of this paper. We define the received power for the $j$-th user in the $i$-th sub-channel is given by
\begin{align}\label{received_power}
P^r_{i,j}(t) = P_{i,j}(t)\mathcal{P}_L(r_{i,j}(t)) |h_{i,j}(t)|^2.
\end{align}

If the receive power obeys $P^r_{i,1}(t) \ge P^r_{i,2}(t) \ge ... \ge  P^r_{i,N_i}(t)$, the decoding order should be from the $1$-st user to the $N_i$-th user. The signal-to-interference-plus-noise ratio (SINR) for the $j$-th user can be expressed as
\begin{align}
\gamma_{i,j}(t) = \frac{P^r_{i,j}(t)}{\sum_{j'=j+1}^{N_i}P^r_{i,j'}(t)+n_0^2}.
\end{align}
Whereas, the SINR of the last user $N_i$ can be written as
\begin{align}
\gamma_{i,{N_i}}(t) = \frac{P^r_{i,{N_i}}(t)}{n_0^2}.
\end{align}

To guarantee the SIC process, a desired data rate $R_{th}$ for all users is required, so the SINR for the $(j-1)$-th user should obeys $B_s\log_2(1+\gamma_{i,j-1}(t)) \ge R_{th}$, otherwise the BS cannot decode information of the $j$-th user.

\subsection{Layer-based Transmit Power Pool}
\begin{figure}
	\centering
	\subfigure[Prototype power pool with corresponding layers]{\label{pool}\includegraphics[scale=.7,keepaspectratio]{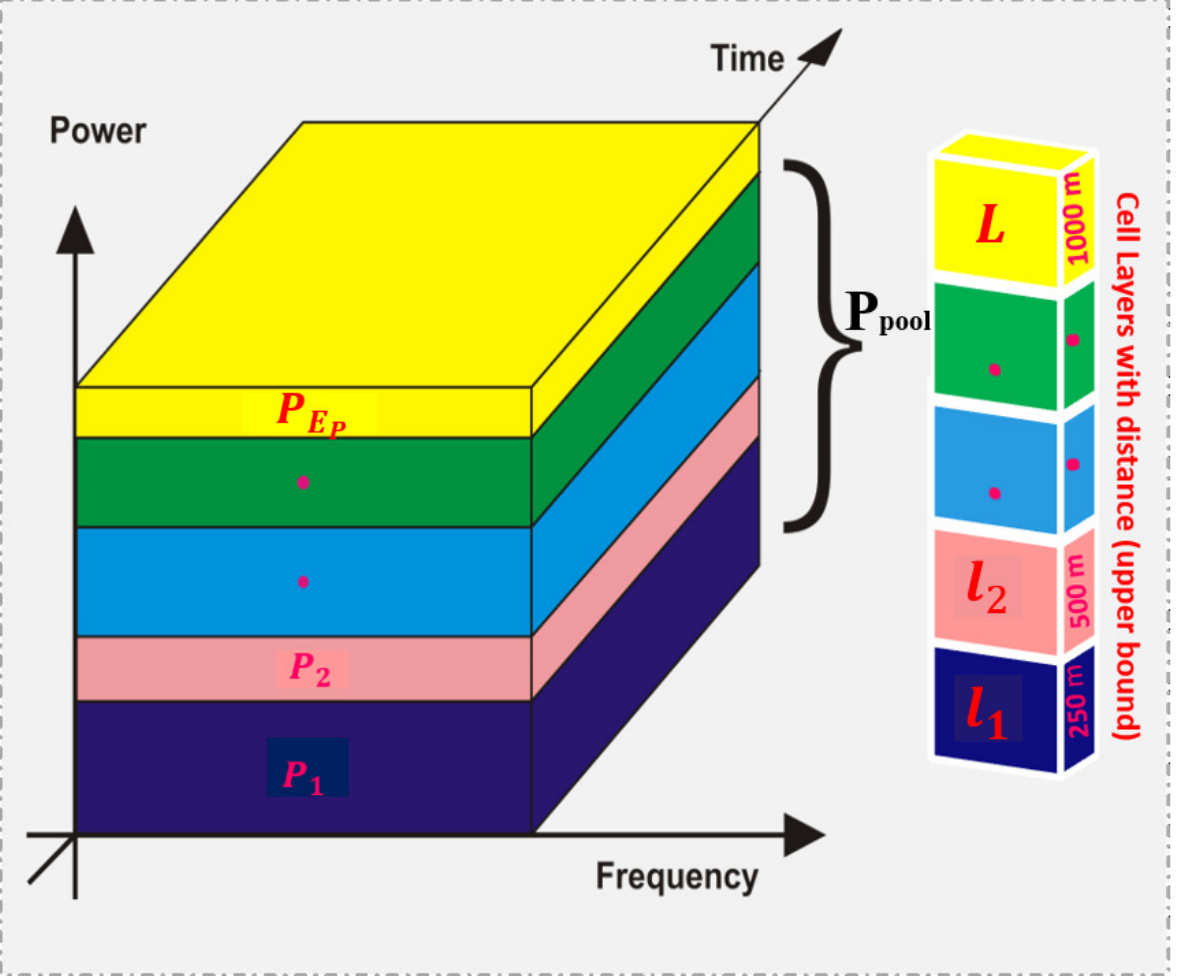}}
	\subfigure[IoT user's procedure flow chart]{\label{pro}\includegraphics[scale=.7,keepaspectratio]{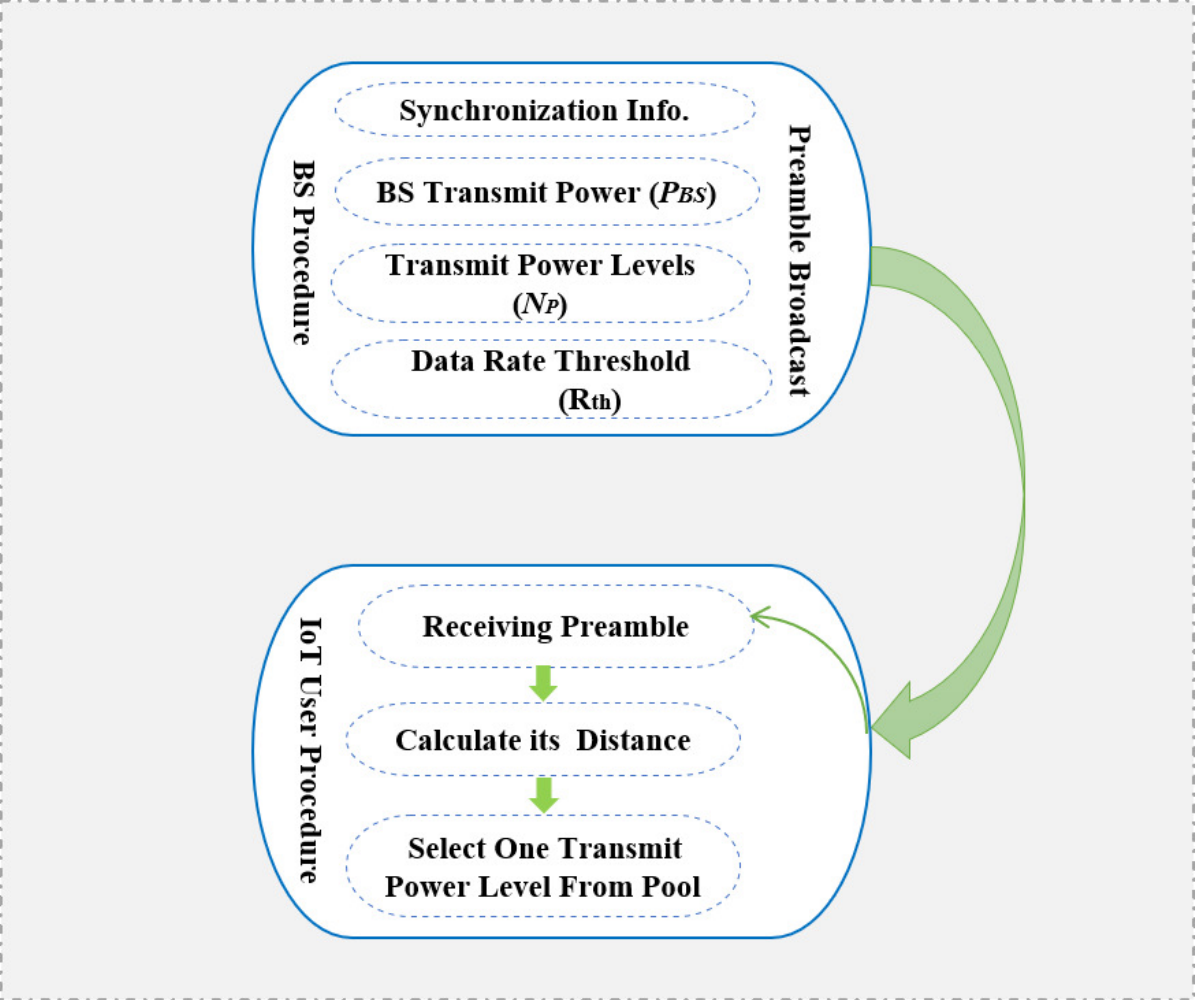}}
	\caption{The designed prototype power pool and working procedure of the BS and IoT users: Sub-figure (a) is the transmit power pool with different power levels for each layer. Sub-figure (b) shows the procedure flow chart between the BS and IoT users.}
	\label{fig:img2}
	\vspace{-0.5 cm}
\end{figure}

From the received power equation \eqref{received_power}, only the transmit power $P_{i,j}(t)$ can be controlled at the user side. Therefore, this paper focuses on designing a prototype of transmit power pool $\mathbf P_{pool} =\{P_1, P_2,...,P_{E_p}\}$ as shown in Fig.~\ref{fig:img2}\subref{pool}, where $\mathbf P_{pool} \subset \mathbf P_t$ and $\mathbf{P}_t = \{P_1, P_2,...,P_{N_p}\}$, $N_p$ is the number of available transmit power levels for IoT users and $E_p$ is the number of power levels in prototype power pool. The set of users $\boldsymbol{V_l}$ in the layer $l$ select one transmit power level from $\mathbf{P}_t$ for uplink transmission.
\begin{remark}
 Due to the fact that NOMA is sensitive to distance-dependent path loss, users can select their transmit power form this prototype power pool solely according to their communication distances to boost up GF-NOMA connectivity efficiency. Moreover, this prototype can be finished offline based on practical stochastic models. To increase the power pool accuracy for any certain application, a further online learning can be employed with the aid of a small load of information exchange.
 \end{remark}

    Without a closed-loop power control, a global resource management in GF-NOMA is a challenging task, as BSs
	commonly have no/partial information about the active users and their channel state information (CSI). Because, with  no/partial information,  keeping enough power difference for SIC processes at the NOMA-enabled BS side is a tough practical challenge. Moreover, exchanging such information with IoT users is expensive in terms of energy consumption and  increases complexity at the BS. Therefore, to avoid acquiring the users' CSI, we assume the BS broadcasts a preamble (which includes transmit power from the BS $P_{BS}$, data rate threshold $R_{th}$, and the transmit power pool $\mathbf{P}_{pool}$) to all users at the beginning of a time slot to synchronize uplink transmissions \cite{8085106}. The working procedure is given in Fig.~\ref{fig:img2}\subref{pro}. In Addition, for offline training (power pool design), we can obtain the location information of all users via GPS, channel estimation or sensing mechanism. For online training (power pool enhancement), we may use the traditional channel estimation method to acquire the CSI.

\subsection{Sub-Channel Selection}
As NOMA provides the opportunity to multiplexed multiple users on the same resource block (RB).
Let $N_{i}$ is the set of users sharing the $i$-{th} sub-channel. To form a NOMA cluster, the condition $|N_{i}|>1$ must be satisfied. Furthermore, we assume that each IoT user is permitted to select at most one sub-channel. For a random IoT user $j$ at time slot $t$, we define a sub-channel selection variable as follows:
\begin{align}\label{channel}
k_{i,j}{(t)}=
\begin{cases}
\text {1}, &\text{IoT user $j$ select sub-channel $i$}\; \\
\text{0}, & \text{otherwise}
\end{cases}
\end{align}

\subsection{Problem Formulation}
To determine the optimal transmit power levels for the design of prototype power pool, we formulate the sum rate maximization as an optimization problem in this section. Therefore, maximizing the sum rate by optimizing the transmit power and sub-channel selection for each user can contribute to the design of a transmit power pool that can be further allocated to a geographically distributed region. More specifically, two matrices $\mathbf{P}_t$ and $\mathbf{K}_t$  need to be optimized to maximize the long-term sum rate. In each time slot $t$, the data rate of an IoT user $j$ over sub-channel $i$ can be written as
\begin{align}\label{Data_Rate}
R_{i,j}(t)=B_s\log_2(1+\gamma_{i,j}(t))
\end{align}
Therefore, the optimization problem can be formulated as
\begin{subequations}\label{of}
	\begin{align}
	\max_{p_{i,j}\in \mathbf{P}_t, k_{i,j}\in \mathbf{K}_t}\sum_{t=1}^{T} \sum_{i=1}^{M} \sum_{j=1}^{N_i}& R_{i,j}(t), \tag{\ref{of}}\\
	\textrm{s.t.}\quad
	P^r_{i,1}(t) \ge P^r_{i,2}(t) \ge &... \ge  P^r_{i,N_i}(t), ~\forall i,\forall t, \label{of0}\\
	\sum_{j=1}^{N_i}P_{i,j}(t)  \leq P_{max}&,\hspace*{0.4in}\forall i, \forall t,\label{of1}\\
	\sum_{i=1}^{M} k_{i,j}(t) \leq 1,~~~~&\hspace*{0.48in}\forall j, \forall t, \label{of2}\\
	N_{i}(t) \geq 2, ~~~~~~~~~~& \hspace*{0.5in}\forall i, \forall t, \label{of3}\\
	\sum_{i=1}^{M}R_{i,j}(t) \geq R_{min}&(t), \hspace*{0.3in}\forall j, \forall t, \label{of4}\\
    E_p < N_p,\hspace*{0.54in}&  \label{of5}\\
	E_p  = L,\hspace*{0.6in}& \label{of6}
	\end{align}
\end{subequations}
where \eqref{of0} is to ensure SIC decoding order, \eqref{of1} represents the maximum power limit $P_{max}$ of each user, in \eqref{of2} the variable $k_{i,j}$ indicates that at each time slot a user $j$ is able to select only one sub-channel, \eqref{of3} represents the minimum number of users on each sub-channel, \eqref{of4} guaranteeing the minimum required data rate $R_{min}$ of each IoT user, \eqref{of5} represents the number of power levels in the prototype power pool $\mathbf{P}_{pool}$ (optimal power level for each layer), and \eqref{of6} represents power levels in the pool should be equal to the number of layers in cell area.
\section{Multi-Agent DRL Based Power Pool Design }
\subsection{Overview of Multi-Agent Deep Reinforcement Learning}
MARL is the extension of single agent RL which involves a set of agents, $G=\{1, 2, 3\cdots \cdots, N\}$, where the whole team of agents acting autonomously and concurrently in a shared environment. MARL can be classified into two cases: MARLs with centralized or decentralized reward. In MARL with centralized rewards, all agents receive a common (central) reward, on the other hand in MARL with decentralized, every agent obtains a distinct reward \cite{lee2020optimization}. However, in multi-agent environment, all agents under decentralized way may compete with each other, i.e., agents may act in a selfish behaviour for requiring the highest reward which may effect the global network performance. To convert this selfishness into a cooperative behaviour, the same reward may be assign to all agents \cite{liang2019spectrum}.  In next section, we apply MARL with centralized reward only to prevent selfish behaviour of agents.

In MARL with centralized reward setting, a multi-agent Markov decision process (MDP) can be represented by a tuple of $(\{S_j\}_{j=1}^N, \{A_j\}_{j=1}^N, P, r )$. Each agent $j$ observes a state $s_j$ from the environment and executes an independent action $a_j$ from its own set of actions $A_j$ on the basis of its local policy  $\pi_j:S_j\rightarrow A_j$. Agents perform joint action $a$ $(a=a_1, a_2, \cdots, a_N \in A)$, where $A =(A_1\times A_2 \times \cdots \times A_N)$, the environment moves from state $s_j^{(t)} \in S_j$ to a new state $s_j^{(t+1)} \in S_j$ with probability of $Pr(s_j^{(t)}|s_j^{(t+1)}, a_j)$, then the agent $j \in G$ receives a common reward $r^{(t+1)}$. Every agent forms an experience $e_j^{(t+1)} = (s_j^{(t)}, a_j^{(t)}, r^{(t+1)}, s_j^{(t+1)})$ at time $(t+1)$, which defines an interaction with the environment \cite{mnih2015human}. The goal of each agent is to learn a local optimal policy $\pi_j^*$ that forms a central optimal policy $\pi^*$ i.e. $(\pi_1^*, \pi_2^*, \cdots, \pi_N^*)=:\pi^*$ for maximizing long term reward \cite{liang2019spectrum}.

We model the selection of transmit power levels and sub-channels in GF-NOMA IoT networks as MDP problem consisting of states $s_j^{(t)} \in S_j$, actions $a_j^{(t)}\in A_j$, and reward $r^{(t)}$ following a policy $\pi_j$. The main elements of the multi-agent DRL based GF-NOMA transmit power pool design are given as follows:
\begin{itemize}
\item State space $\boldsymbol{S}$: To explore environment feature, each IoT user $j$ acts as an agent and simultaneously interacts with unknown environment. We define data rate of IoT users as the current state $s_j^{(t)} \in S_j$, where
\begin{align}\label{state}
S_j= \{R_{1,1}^{(t)},R_{2,1}^{(t)},\dotsi R_{i,j}^{(t)},\dotsi,R_{M,N}^{(t)}\},
\end{align}
and $R_{i,j}$ is the data rate of user $j$ on sub-channel $i$ in time slot $t$.  Moreover, the state size is equal to the number of active IoT users $N_t$ in a time slot $t$.
\item Action space $\boldsymbol{A}$: Action $a_j^{(t)} \in A_j$ of agent $j \in G$ is the selection of power level $p\in P$ and sub-channel $m \in M$. The transmission power is discretized into $N_p$ power levels, hence the dimension of action space is $N_p\times M$, where $M$ is the number of sub-channels. The action space is given by
\begin{align}\label{action}
A =(A_1\times A_2 \times \cdots \times A_j \cdots \times A_N),\\
\text{where}~A_j= \{1,2,\cdot\cdot\cdot pm,\cdot\cdot\cdot,P_{N_p}M\}.
\end{align}
If an agent (IoT user) $j$ transmits with power level $p$ on sub-channel $m$ in time slot $t$, then the corresponding action is $a_j^{(t)}\in A_j=pm$, i.e., each action corresponds to a particular combination of power level and sub-channel selection.
\item Reward $\boldsymbol{Re}$: The system performance depends on reward function flexibility and its correlation with the objective function \cite{liang2019spectrum}. To enhance system performance we represent sum throughput of GF-NOMA system as a reward signal, which is strongly correlates with the objective function. An agent $j$ receives a returned reward $r^{(t)} \in Re$  after choosing action $a_j^{(t)}$ in state $s_j^{(t)}$ in a TS $t$ determined by
\begin{align}\label{Reward}
r^{(t)} = \sum_{i=1}^{M} \sum_{j=1}^{N_{t,i}} R_{i,j}.
\end{align}
In our proposed model the short term reward of an agent $j$ depends on the following conditions
  \begin{multline}\label{Re}
    r_j^{(t)}=
    \begin{cases}
    \text {$r^{(t)}$}, \text{ if $R_{current}\geq R_{previous}$ and } \\ \hspace*{0.3in}\text{satisfying constraints given in}
    \\ \hspace*{0.3in}\text{ (\ref{of0})-(\ref{of4})} \; \\
    \text{0},  \hspace*{0.2in}\text{otherwise}.
    \end{cases}
  \end{multline}
This reward or penalty can help agents to find optimal actions that can maximize cumulative reward for all interactions with the environment.
\end{itemize}
Classic Q-Learning algorithm \cite{sutton2018reinforcement}, aims to compute an optimal policy $\pi^*$ by maximizing expected reward. The long term discounted cumulative reward at time slot $t$ is given by

\begin{align}\label{Reward1}
Re^{(t)} = \sum_{k=0}^{\infty }\gamma^k r^{(t+k+1)},
\end{align}
where $\gamma \in [0,1]$ is the discount factor.
Q-Learning is based on action-value function, the Q-function for IoT agent $j$ which is defined as the expected reward after taking action $a_j$ in state $s_j$ following a certain policy $\pi$ \cite{singh2000convergence}, can be expressed as
\begin{align}\label{Q-func}
Q_j^\pi(s_j, a_j) = \mathbb E^\pi\Big[Re^{(t)}\big|s_j^{(t)}=s, a_j^{(t)}=a\Big],
\end{align}
where corresponding values of (\ref{Q-func}) is known as action values or Q-values and satisfies a Bellman equation,
\begin{multline}\label{bellman}
Q_j^\pi(s_j, a_j) =R(s_j, a_j) + \\\gamma \sum_{s_j^\prime \in S_j} P_{s_j\rightarrow s_j^\prime}^a \Big( \sum_{a_j\in A_j} \pi (s_j^\prime, a_j^\prime) Q^\pi(s_j^\prime, a_j^\prime)\Big),
\end{multline}
where $R(s_j, a_j)$ is the immediate reward by taking action $a_j$ in state $s_j$ and $P_{s_j\rightarrow s_j^\prime}^{a_j}$ is the transition probability from state $s_j$ to new state $s_j^\prime$ by selecting action $a_j$. By solving MDP each IoT agent is able to find the optimal policy $\pi^*$ to obtain maximal reward. The optimal Q-function for IoT agent $j$ associated with policy $\pi^*$ can be expressed as
\begin{align}\label{bellman}
Q_j^{\pi^*}(s_j, a_j) =R(s_j, a_j) + \gamma \sum_{s_j^\prime \in S_j} P_{s_j\rightarrow s_j^\prime}^{a_j}\max_{a_j^\prime} Q^*(s_j^\prime, a_j^\prime).
\end{align}
The quality of a given action in a state can be measured by its corresponding Q-value.
To maximize its reward and improve policy $\pi$, agent $j$ decides its action from
\begin{align}\label{argmax}
a_j & =\argmax_{a_j\in A_j} \; Q(s_j, a_j).
\end{align}
In Q-learning algorithm, to store Q-values of all possible state-action pairs, every agent needs to maintain a lookup table (Q-table), $q(s_j, a_j)$ as a substitute of optimal Q-function. After random initialization of the Q-table, for each time step all the agents take actions according to  the $\epsilon$-greedy policy. With probability $\epsilon$, all agents decides actions randomly to avoid sticking in non-optimal policy, whereas with probability of $1-\epsilon$, agents select actions that gives maximum Q-values for the given state \cite{mnih2015human}. After taking action $a_j$ in a given state $s_j$, the agents acquire a new experience, and Q-learning algorithm updates its corresponding Q-value in the Q-table.

During the decision process in a time slot $t$,  if an agent $j$,  given a state $s_j^{(t)}$, selecting action $a_j^{(t)}$, receiving a reward $r^{(t)}$ and the next state $s_j^{(t +1)}$, then its associated Q-value is updated as
\begin{align}\label{Qup}
Q(s_j^{(t)}, a_j^{(t)}) &\leftarrow r^{(t)}+\gamma \max_{a_j\in {A}_j} Q(s_j^{(t+1)}, a_j).
\end{align}
However, for IoT scenario, the size of Q-table increases with the increasing number of state-action spaces (an increase of IoT users) that makes Q-learning expensive in terms of memory and computation because of the following two reasons..
\begin{enumerate}
	\item Several states are infrequently visited, and
	\item Q-table storage in \eqref{Qup} becomes unrealistic.
\end{enumerate}
In addition, DRL is one of the RL algorithms, which tends to obtain more rewards as per its efficient learning behaviour, in comparison with the traditional Q-learning algorithm which are prone to negative rewards \cite{ahsan2020resource}.
\begin{figure*}[t!]
	\centering
	\includegraphics[width = 1 \linewidth,keepaspectratio]{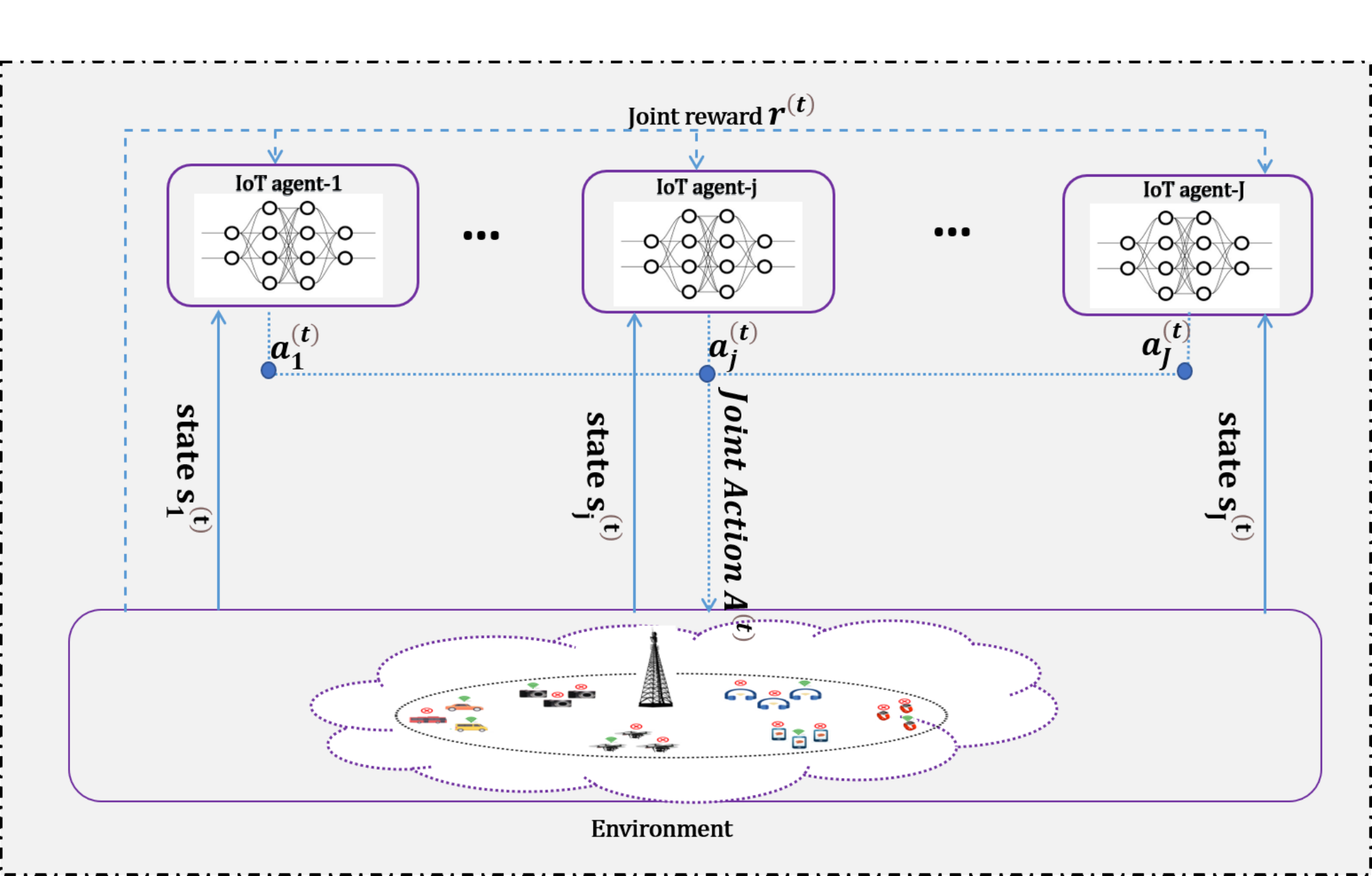}\\
	\caption{The MARL framework for GF-IoT Networks, where IoT agents takes joint actions and receives a common reward during the training process to learn the optimal policy. }\label{nn}
\end{figure*}

To overcome the above problems, DRL (e.g. Deep Q Network algorithm) is proposed, \cite{mnih2015human}, in which the Q learning is combined with Deep Neural Network (DNN) for Q function approximation $Q(s, a; \theta)$, where $\theta$ represents its parameters (weights). Hence keeping a large storage space for state-actions pair (Q-values), DRL agent only memorize $\theta$ weights in its local memory that reduces the memory and computation complexity.

In MARL based DRL setting, each agent $j$ has a DQN that takes the current state $s_j^{(t)}$ as input and output Q-value function of all actions. IoT agents explore the environment by state-action pair following $\epsilon$ greedy policy. Every agent collects and stores the experiences in the form of a tuple $\big(s_j^{(t)}, a_j^{(t)}, r^{(t)}, s_j^{(t+1)}\big)$ in replay memory. In each iteration, a mini-batch of data is sampled uniformly from the memory and is used to update network weights $\theta$. The target value produced by target Q network from randomly sampled tuple $\big(s_j^{(t)}, a_j^{(t)}, r^{(t)}, s_j^{(t+1)}\big)$ is given by
\begin{align}\label{argmax1}
y_j^{(t)} & =r^{(t)}+ \gamma\argmax_{a_j^{(t+1)}\in A_j} \;Q(s_j^{(t+1)}, a_j^{(t+1)}; \bar{\theta}),
\end{align}
where, $\bar{\theta}$ represents the weights of the target Q-network, which are set equal to the weights $\theta$ of the online training network after every $T_U$ steps.
The Q network can be trained by minimizing the loss function by using variant of stochastic gradient decent,
\begin{align}\label{loss1}
L(\theta) & = (y_j^{(t)}-Q_j^{(t)}(s_j^{(t)}, a_j^{(t)}; \theta))^2.
\end{align}
The policy $\pi$ used by each user for selecting power level and sub-channel is random at the start and gradually improved to the optimal policy $\pi^*$ with the updated Q-networks.

\subsection{Proposed Multi-Agent DRL-based Grant-Free NOMA Algorithm}
Designing a model-free distributed learning algorithm for solving an optimization problem that can effectively adapt to topology changes, different objectives, and general complex real-world environments while overcoming expensive computational requirements due to the large state space and partial observability of the problem is a challenging task. Another challenge is enabling a large number of users to share the limited resources in GF transmissions. Where the goal of users is to maximize a given network utility in a distributed manner without sharing information or performing online coordination. Our algorithm and problem definition are fundamentally different from those discussed in Section-I (A).
\begin{figure*}[t!]
	\centering
	\includegraphics[width = 1 \linewidth,keepaspectratio]{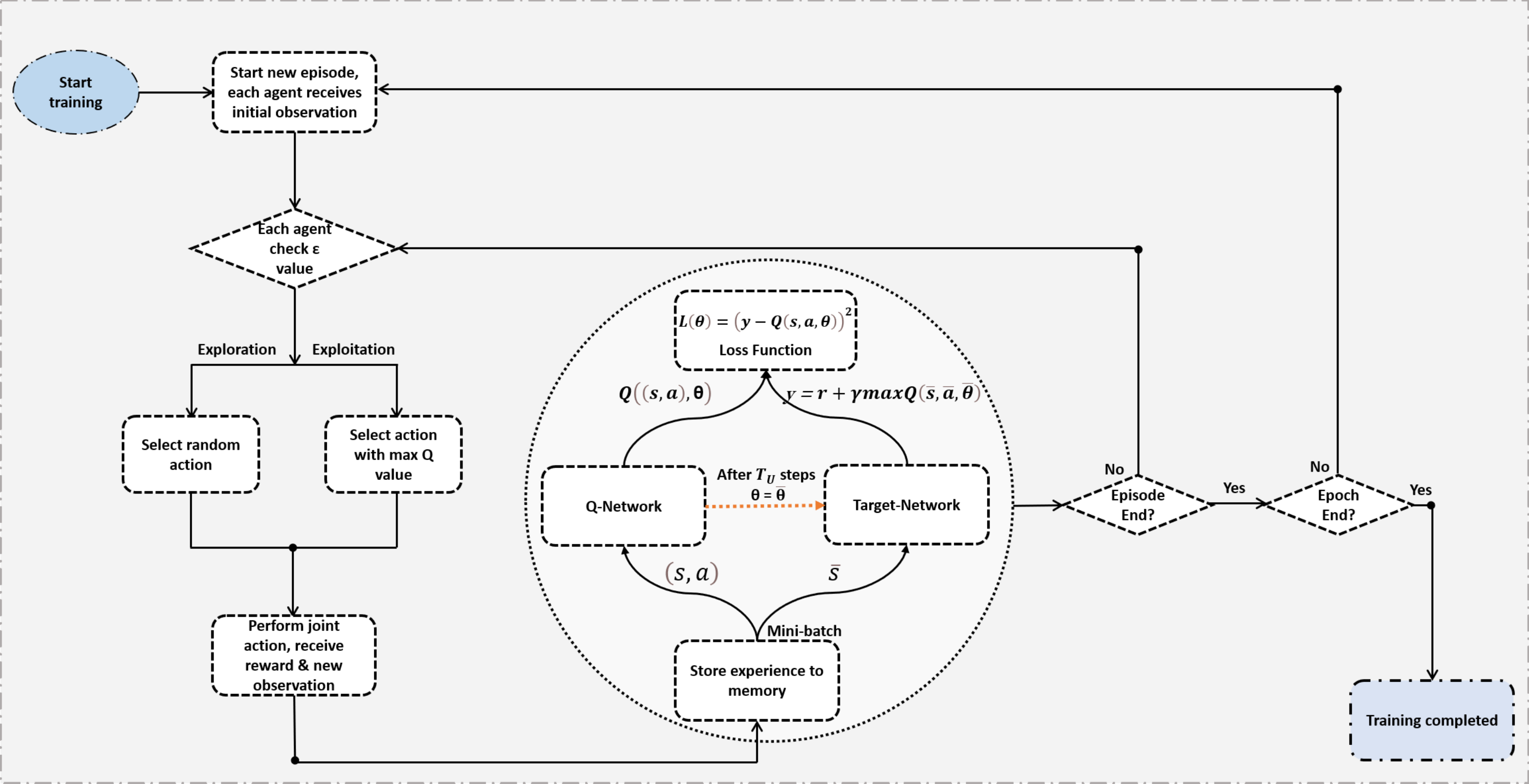}\\
	\caption{The proposed end-to-end architecture for training deep MARL agents to optimize the power and sub-channel allocation. }\label{alg}
\end{figure*}
For example, an optimal solution is a combinatorial optimization problem with partial state observations which is mathematically intractable as the network size increases. Furthermore, our algorithm differs from the existing DRL-based schemes, where the communication is based on closed-loop power control and ACK signals, by eliminating the process of ACK signals and providing open-loop power control, which reduces the computational complexity at the BS. Therefore, our proposed algorithm is more suitable and scalable for IoT based applications regardless of the number of users and their locations in the network.
Next, we describe DQN and its variant DDQN based GF-NOMA algorithm for sub-channel and power allocation as
\begin{algorithm}
	\fontsize{9pt}{12pt}
	\selectfont
	\caption{\small Proposed Multi-Agent DQN-Based Grant-Free NOMA Algorithm}
	\begin{algorithmic}[1]
		\State \textbf {Parameter initialization Phase:}
		\State Initialize parameters $N_p, M, \gamma, \alpha, \epsilon$, batch-size $N_b$, $T_u$
		\State Set replay memory with size $D$
		\State Initialize Q-network weights for all agents (IoT users) and copy primary Q-network weights to Target Q-network
		\State \textbf {Training Phase:}
		\For {$episode=1,2,\ldots, E$}
		\State reset the environment
		\For {$Time~step=1,2,\ldots,F$}
		\For {each IoT agent $j$}
		\State Input state $s(t)$
		\State Take action $a(t)$ based on (\ref{greedy})		
		\EndFor
		\State all agents take joint actions observe new state $s(t+1)$ \hspace*{0.4in}and reward $r(t)$ using (\ref{Re})
		\For {each IoT agent $j$}
		\State Store a tuple of $s(t), a(t), r(t), s(t+1)$ in replay \hspace*{0.6in}memory \nonumber
		\EndFor
		\For {each IoT agent $j$}
		\State Uniformly select batches from memory $D$
		\State using stochastic gradient to minimize loss between the \hspace*{0.6in}primary network and target network:
		\[\Big[y_j^{(t)}-Q_j^{(t)}(s_j^{(t)}, a_j^{(t)}; \theta)\Big]^2\]
		\EndFor
		\If {$  episode\% ==T_u$ }
		\State Update target Q-network weights
		\EndIf
		\EndFor
		\EndFor
	\end{algorithmic}

\end{algorithm}
\subsubsection{Deep Q-Learning}
The schemes based on classic Q-learning are able to work well on small number of states and actions. However, when the states and actions spaces increases, such approaches incorporates many limitations as described in Section III-A. If the problem size and number of agents is sufficiently large, convergence rate might be slow and cannot be implemented in real time scenarios. DRL promotes and inherits the advantages of deep learning and RL techniques that overcome the problems of classic Q learning.

As depicted in Fig.~\ref{nn}, we proposed an independent learners (IL) based MARL algorithm to solve the optimization problem mentioned in \eqref{of} in-order to design transmit power pool. In the proposed multi-agent DRL-based GF NOMA algorithm shown in \textbf{Algorithm 1}, each IoT user runs an independent DQN and jointly learns an individual optimal strategy for solving the formulated MDP.
\begin{remark}
In multi-agent system, the Q-function of each agent is based on the joint actions and joint strategy due to the reason that finding an optimal strategy is difficult \cite{neto2005single}. To encounter this problem, each IoT user acts as an IL and ignores actions and rewards of other users and interacts with the environment in such a way that no other IoT user exists\cite{cui2019multi}\cite{9205989}.
\end{remark}
\begin{algorithm}
	\fontsize{9.2pt}{12pt}
	\selectfont
	\caption{\small Proposed Multi-Agent Double DQN-Based GF-NOMA Algorithm}
	\begin{algorithmic}[1]
		\State \textbf {Parameter initialization Phase:}
		\State Initialize parameters $N_p, M, \gamma, \alpha, \epsilon$, batch-size $N_b$, $T_u$
		\State Set replay memory with size $D$
		\State Initialize Q-network weights for all agents (IoT users) and copy primary Q-network weights to Target Q-network
		\State \textbf {Training Phase:}
		\For {$episode=1,2,\ldots, E$}
		\State reset the environment
		\For {$Time~step=1,2,\ldots,F$}
		\For {each IoT agent $j$}
		\State Input state $s(t)$
		\State Take action $a(t)$ based on (\ref{greedy})		
		\EndFor
		\State all agents take joint actions observe new state $s(t+1)$ \hspace*{0.4in}and reward $r(t)$ using (\ref{Re})
		\For {each IoT agent $j$}
		\State Store a tuple of $s(t), a(t), r(t), s(t+1)$ in replay \hspace*{0.6in}memory \nonumber
		\EndFor
		\For {each IoT agent $j$}
		\State Uniformly select batches from memory $D$
		\State using stochastic gradient to minimize loss between  \hspace*{0.6in}the primary network and target network:
		\[\Big[r^{(t)}+ \gamma Q(s_j^{(t+1)}, \argmax_{a_j^{(t+1)}\in A_j} \;Q(s_j^{(t+1)}, a_j^{(t+1)}))-Q_j^{(t)}(s_j^{(t)}, a_j^{(t)}; \theta)\Big]^2 \]
		\EndFor
		\If {$  episode\% ==T_u$ }
		\State Update target Q-network weights
		\EndIf
		\EndFor
		\EndFor
	\end{algorithmic}
\end{algorithm}
The training procedure of our proposed algorithm is depicted in Fig.~\ref{alg}. The learning parameters i.e., learning rate $\alpha$, discount factor $\gamma$, memory size $D$, batch size $N_b$, and $\epsilon$ are initialized and configured during initialization phase. While in training phase, through interaction with environment the agents initially collects training samples. In a time slot $t$, each agent $j$ inputs current state $s_j^{(t)}$ into primary Q-network and obtains all the Q-values corresponding to all actions. The action $a_j^{(t)}$ is determined by the following policy,
\begin{align}\label{greedy}
a_j^{(t)}=
\begin{cases}
\text {random action},& \text{with probability}\; \epsilon, \\
\argmax\limits_{a_j^{(t)}\in A_j} \;Q(s_j^{(t)}, a_j^{(t)}) & \text{with probability}\;1-\epsilon.
\end{cases}
\end{align}
To fully explore the environment and to find the action that returns the best reward for a given state random action (try and observe) strategy with small probability $\epsilon$ is considered. All agents perform a joint action based on the above policy, receives a common reward and the environment transitions to a new state $s_j^{(t+1)}$. Thus, each agent make a new training sample, and stores it in memory $D$.

We sample batches of stored transitions from the replay memory and compute target Q-value of each sample. In each training step, with target Q-value and selected samples, the primary Q network is trained through minimizing the error with gradient descent method using (\ref{loss1}) to improve policy. After $T_u$ training steps, target Q-network weights are sets as the weights of the primary Q-network. The training process is finished either after reaching total training steps or a predefined number of episodes.

\subsubsection{Double Deep Q-Learning}
As traditional DQN tends to significantly overestimate action-value that leads to poor policy and unstable training. The accuracy of Q values depends on the actions that agent takes and states explored, however, at the beginning of training the agent don't have sufficient information about the optimal action to take. Therefore, considering the maximum $q$ value (may be noisy) as the optimal action to take can lead to maximization bias in learning. Instead of best optimal actions, if non-optimal actions are given higher Q values, then learning will be unstable and complicated. Moreover, in DQN, the max operation uses the same values for both selection and evaluation of an action. This can, therefore, lead to overoptimistic value estimates \cite{nguyen2019non}. For example, if action $a$ has higher value than action $b$ in some state $s$, then agent will choose action $a$ every time for that state $s$. Now suppose if for some memory experience action $b$ becomes the better action then since the neural network is trained in a way to give a much higher value for action $a$, it is difficult to train the network to learn that action $b$ is better than action $a$ in some conditions. To bring down the difference between the output values (actions) and to mitigate the problem of overoptimistic and biased value estimation, DDQN is proposed\cite{van2015deep}. The idea behind DDQN algorithm is to prevent this bias estimation by decoupling the max operation in the target into action selection and action evaluation process. More specifically, the agent uses
\begin{itemize}
	\item DQN network to select the optimal action (action with highest Q value) for the next state.
	\item The target network to compute the target Q value of the corresponding action in next state.
\end{itemize}
\begin{remark}
We use a secondary model that copy the main model from the last episode, because the difference between values of the secondary model is less than the main model, thus we use this second model to obtain the Q values.
\end{remark}

The target of DDQN can be written as:
\begin{align}\label{DDQNV}
y_j^{(t)} & =r^{(t)}+ \gamma Q(s_j^{(t+1)}, \argmax_{a_j^{(t+1)}\in A_j} \;Q(s_j^{(t+1)}, a_j^{(t+1)}; \theta); \bar{\theta}).
\end{align}
DDQN updates is the same as traditional DQN, however, replace the $y_j^{(t)}$ with (\ref{DDQNV}). Details of our proposed multi-agent DDQN GF-NOMA algorithm is given in \textbf{Algorithm 2}.
\subsection{Computational Complexity}
The number of operations using the DQN with $H$
layers, in which $Z$ is the input layer size and is proportional
to the number of active IoT users $N_t$ in the network, and $x_h$ is the number of neurons in layer $h$. These parameters obey $X\triangleq Zx_1 + \sum_{h=1}^{H-1}x_hx_{h+1}$. Thus,  at each time step, real-time computational complexity for each IoT user is given by $\mathcal{O}(X)$. For a single sample, the computational complexity of the forward and back propagation is  $\mathcal{O}(X)$  and the training complexity
for $ N_t$ users (agents), one minibatch of $E$ episodes with $F$ time-steps until convergence results in computational complexity of order $\mathcal{O}(EFN_tX)$ in training phase. To find the optimal solution using exhaustive search, we need to consider all possible combinations of power and sub-channel selection for all users, which is not feasible due to the large state and action spaces of our problem. Furthermore, in case of mobile users, there might be a limited impact on the proposed algorithm. However, when the speed of users is fast, the correlation between different layers becomes high due to Doppler effects and multi-path delays. Then some adjustments based on the proposed solution are required.
\section{Experimental Results}
\subsection{Simulation Setup and System Parameters}
In this section, simulation results are presented to demonstrate the proposed algorithm performance. We consider a single BS and users are activated following a Poisson distributed traffic. All active users communicate with the BS in a GF manner with in a radius of $1000 m$. The entire area of the cell is divided into $4$ layers, each of which is $250m$ wide. Other simulation parameters are summarized in Table~\ref{tab1}. We evaluate the performance of our proposed algorithm on Intel core i5-8265 CPU with 1.8 GHz frequency with 8GB of Random Access Memory and 64-bit operating system (Windows 10). We use a deep neural network for DQN with fully connected hidden layers and Rectified Linear Unit (ReLU) as the activation function for hidden layers. The Q-network input layer size (i.e, state size) is equal to the number of users in the network. DQN output layer size is equal to the number of all possible actions, i.e., $Np \times M$. We utilize the $\epsilon$- greedy policy to balance exploration and exploitation phenomena. Other training parameters are given in Table \ref{tab2}. Tuning some of these parameters can improve the performance of deep neural networks, but the one that could accelerate algorithm converging or exploding is the selection of optimizer. We choose the Adam optimizer \cite{kingma2014adam}, as it restricts the oscillations in vertical direction and the learning rate can be increased to take larger steps for fast convergence.
\begin{table}[t!]
    \footnotesize
	\centering
	\caption{Network Settings of the Proposed System}
	\label{tab1}
	\begin{tabular}{l|l}
		\hline
		\hline
		\textbf{Power levels} & \text{[0.1, 0.2, 0.3, 0.4, 0.5, 0.6, 0.8, 0.9] W} \\ \hline
		\textbf{Path loss exponent} & $\alpha=4$ \cite{zhang2020deep}  \\ \hline
		{\textbf{AWGN($N_0$)}} &{ -90dBm \cite{zhang2020semi}}\\ \hline
		{\textbf{No. of sub-channels}} & {[2, 3, 4]}\\ \hline
		{\textbf{Sub-channel bandwidth}} & {$B_s=10$ KHz \cite{zhang2020deep} }\\  \hline
		{\textbf{Carrier frequency} }& {10 MHz \cite{zhang2020semi}} \\\hline
		{\textbf{Minimum data rate }}  & {10 bps/Hz  \cite{9044821}}\\ \hline
		{\textbf{$P_{max}$}} & {1 W \cite{9044821}} \\	
		\hline
		\hline
	\end{tabular}
\end{table}

\begin{table}[t!]
	\scriptsize
	\centering
	\caption{DQN Training Parameters}
	\label{tab2}
	\begin{tabular}{l|l}
			
			\hline
			\hline
			\textbf{No. of episodes} & 500  \\ \hline
			\textbf{Layers} & \{Input, hidden layer1, hidden layer2, hidden layer3, output\} \\ \hline
			\textbf{Neurons per layer} & \{250, 120, 64\}  \\ \hline
			\textbf{$\epsilon$} & 1.0                      \\ \hline
			\textbf{$\epsilon$ min} & 0.01                  \\ \hline
			\textbf{Learning rate} & 0.001   				   \\\hline
			\textbf{Optimizer} & Adam 					         \\
			\hline
			\hline
			
	\end{tabular}
\end{table}

\subsection{Multi-Agent DRL-based GF-NOMA Algorithms Performance Analysis}
\begin{figure}[t!]
	\centering
	\subfigure[Convergence comparison]{\label{convergence}\includegraphics[scale=.6,keepaspectratio]{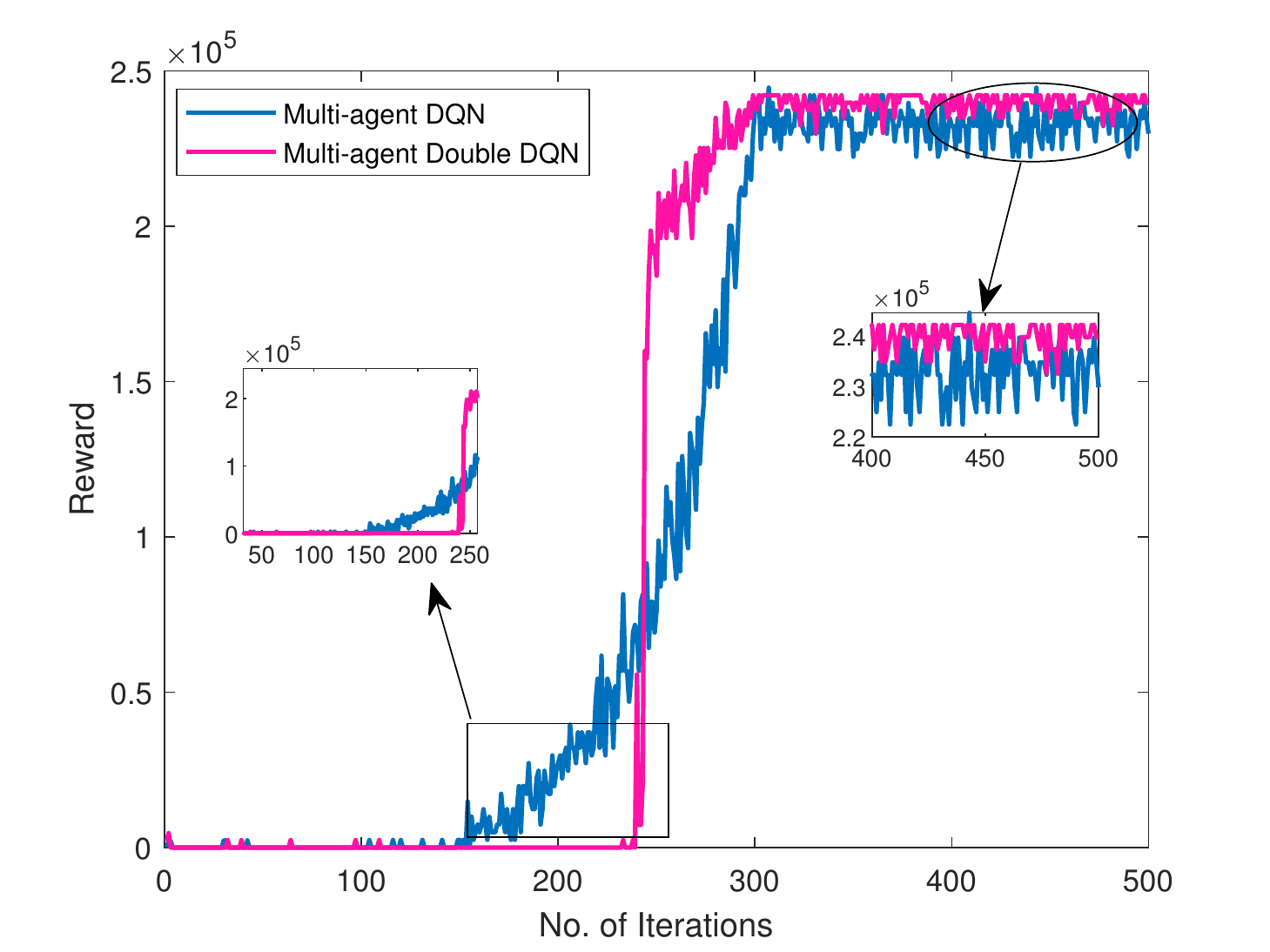}}
	\subfigure[Loss in training phase]{\label{loss}\includegraphics[scale=.6,keepaspectratio]{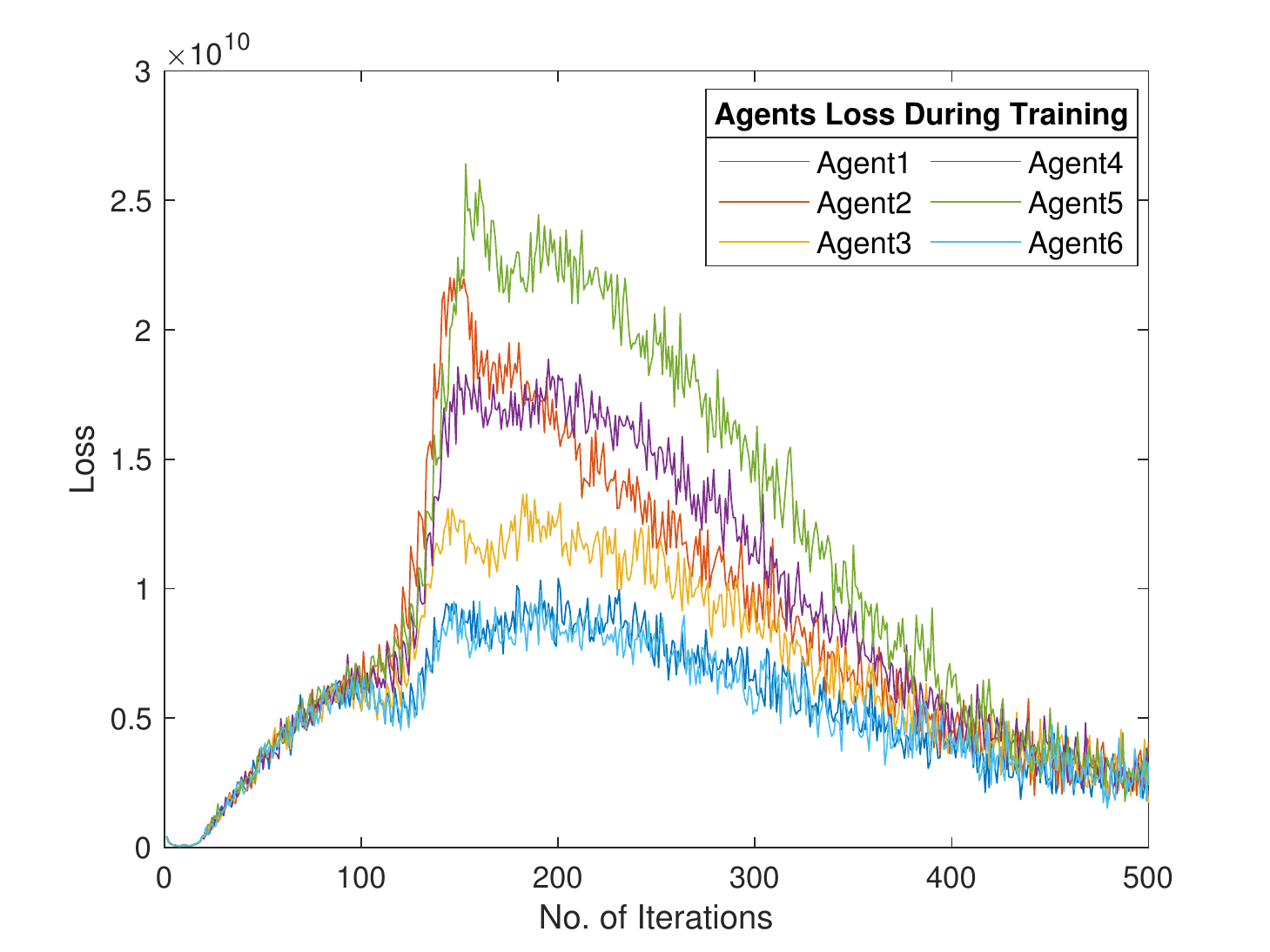}}
	\caption{The convergence of multi-agent DRL based GF-NOMA algorithms and the loss in the training phase: Sub-figure (a) shows the convergence of the proposed multi-agent DQN based GF-NOMA vs. multi-agent DDQN based GF-NOMA algorithm. Sub-figure (b) represents loss function value of all agents ($N_t=6$) in the training phase. At each time step, the stochastic gradient descent algorithm minimizes the loss function mentioned in (\ref{loss1}) for training the mini-batch.}
	\label{cell_area}
	\vspace{-0.5 cm}
\end{figure}

As the convergence of algorithms affects the performance of systems, we analyze and compare the convergence of both algorithms (multi-agent DQN based GF-NOMA and multi-agent DDQN based GF-NOMA) given in Fig.~\ref{cell_area}\subref{convergence}. A clear performance difference between the two algorithms is evident in terms of convergence and the score achieved by the agents. It can be seen that after 300 iterations, both algorithms converge to the optimal value. However, it is observable that multi-agent DDQN based GF-NOMA algorithm obtained a reward (throughput) greater than the traditional DQN based GF-NOMA algorithm. At the beginning of the training, the performance of both algorithms is worst due to exploration phenomena (random actions). After 150 episodes approximately, the agents gain a lot of experience and start to exploit better actions. DQN based GF-NOMA algorithm improves its policy and gradually increases the reward and converges almost in the next 150 episodes. Similarly, agents in multi-agent DDQN based GF-NOMA algorithm gets a reward of zero in about the first 220 episodes. However, there is a sudden surge in reward from zero to $2\times10^5$ in the next few episodes and then a gradual rise in reward for about 25 episodes. Multi-agent DDQN based GF-NOMA algorithm converges for the last 200 episodes. It can be seen that learning with the DDQN algorithm is more stable and performs well as compared to the DQN algorithm. DQN algorithm suffers from the problem of Q values overestimation which leads a low-quality policy.
In Fig.~\ref{cell_area}\subref{loss}, the loss function values of the proposed multi-agent DDQN based GF-NOMA algorithm during the training are shown. The loss function values of all agents reach its peak in about 200 episodes and then decreases as the agents exploit better actions. As the reward converges, the loss value of all agents continuously declines and reaches to its minimum at the end of episodes that justifies the accurate Q values predication. In our proposed multi-agent DDQN based GF-NOMA algorithm, agents are ILs who try is to maximize their rewards. Every agent takes actions independently which may affect other agents performance, therefore we design the reward function in such a way that all users receive the same reward which allows the agents to select those actions that increase the cumulative reward. Hence, every agent updates its policy with the action of other agents.
\subsection{Impact of Learning Rate on Double DQN Performance}
\begin{figure}[t!]
	\centering
	\subfigure[Performance with different learning rate]{\label{lr}\includegraphics[scale=.6,keepaspectratio]{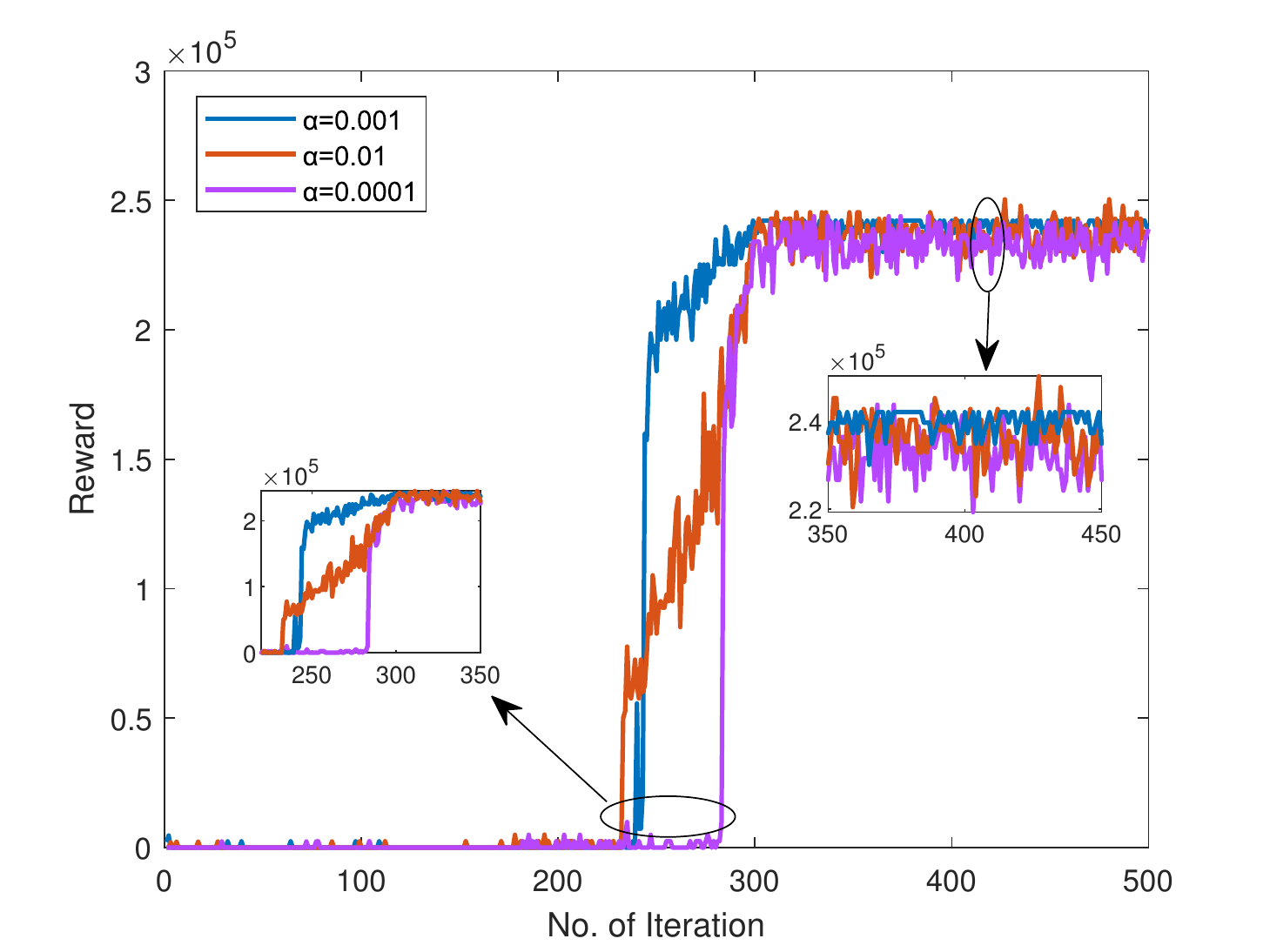}}
	\subfigure[Transmit power constraint impact on system throughput]{\label{power}\includegraphics[scale=.6,keepaspectratio]{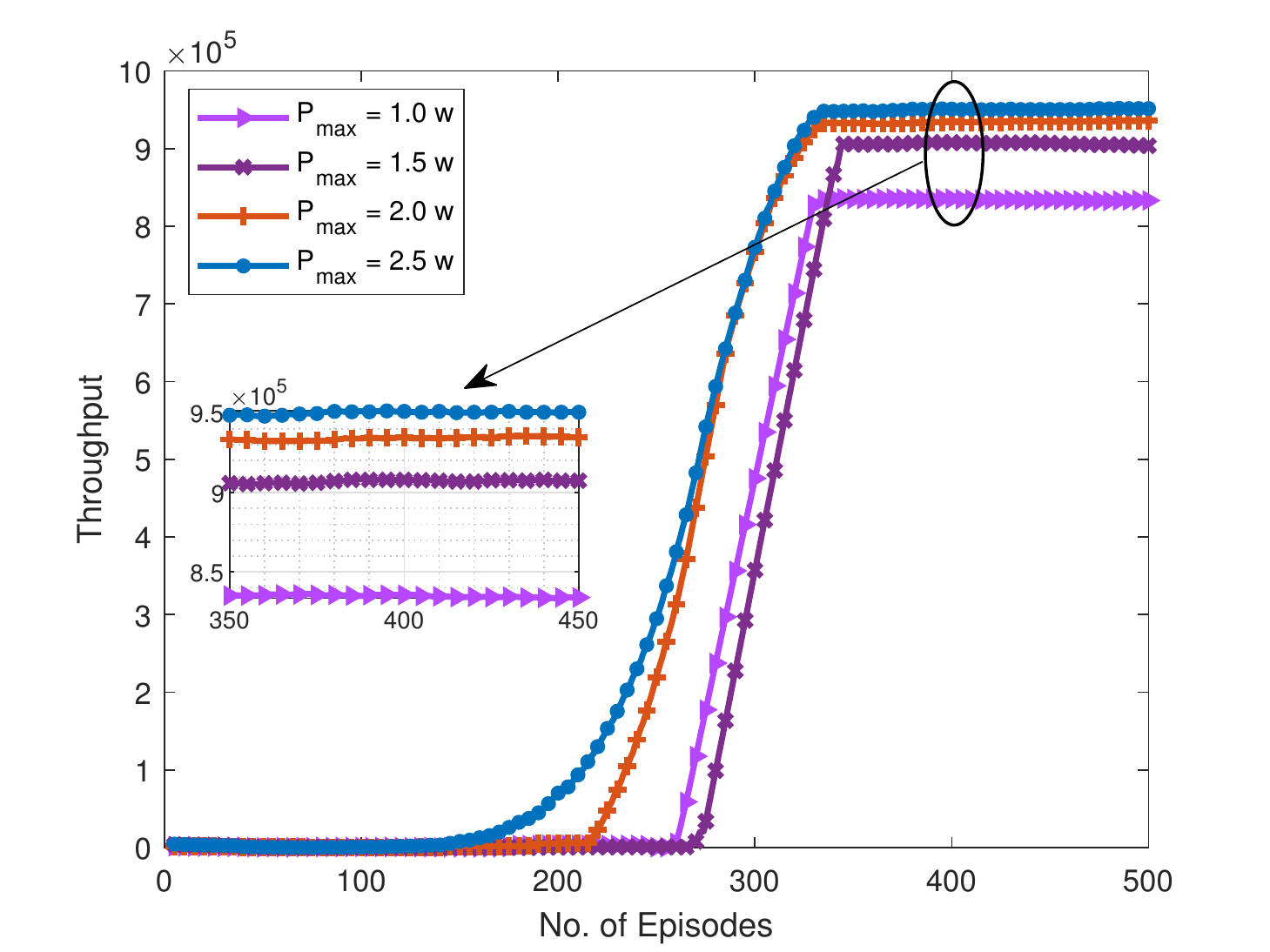}}
	\caption{The learning rate comparison and impact of transmit power constraint: Sub-figure (a) represents the performance comparison with different learning rates. Sub-figure (b) shows the impact of transmit power constraint on the system throughput.}
	\label{LR_loss}
	\vspace{-0.5 cm}
\end{figure}

We use stochastic gradient descent (SGD), an optimization algorithm, to train the deep neural networks. SGD evaluates the error gradient of the current state of the model using some samples from the training data set and then updates the weights of the model using back-propagation. The amount at which the weights are updated is known as learning rate or step size that has a  value in the range of 0.0 and 1.0. The learning rate is a configurable hyper-parameter that controls how fast the model adapts to the problem or how quickly DQN learns from the data. The most important hyper-parameter is the learning rate, if you have time for tuning only one parameter, tune the learning rate \cite{goodfellow2016deep}. Selecting the optimal learning rate is a challenging task as a small learning rate may require large training time due to the smaller changes made to the weights in each update. Whereas, large learning rate value reduces training time due to rapid changes in the weights but the model may converge too fast to a suboptimal solution. Fig.~\ref{LR_loss}\subref{lr} shows the throughput (reward) vs. number of episodes with different learning rates. It can be seen that a large value of learning rate results in fast convergence with large fluctuation in reward value that may lead to unstable training or even to diverge. On the other hand, the too small value of learning rate takes a lot of time to find the optimal policy. The moderate learning rate value of $\alpha=0.001$ shows the highest and more stable reward in terms of throughput. Hence, we opted and set $\alpha=0.001$ as the learning rate value for our simulations.
\subsection{Impact of the Power Constraint}
Fig.~\ref{LR_loss}\subref{power} shows the system performance in terms of throughput with different transmission power constraints. It can be seen that the proposed algorithm achieves the highest throughput at large $P_{max}$ compared to the other values. This is because all the users in the network are able to transmit with large transmit power. In contrast, this constraint with a small value restricts the users to use a small transmit power for uplink transmission and hence a lower throughput is reported. Moreover, a small difference of throughput is presented when $P_{max}= 2.0$ and $P_{max}= 2.5$. It is concluded that a large value of $P_{max}$ does not always contribute to high system performance, since users transmitting at high power increase the inter user interference .

\begin{figure}[t!]
	\centering
	\subfigure[Spectral efficiency vs. No. of Sub-channels]{\label{fig:7(a)}\includegraphics[scale=.6,keepaspectratio]{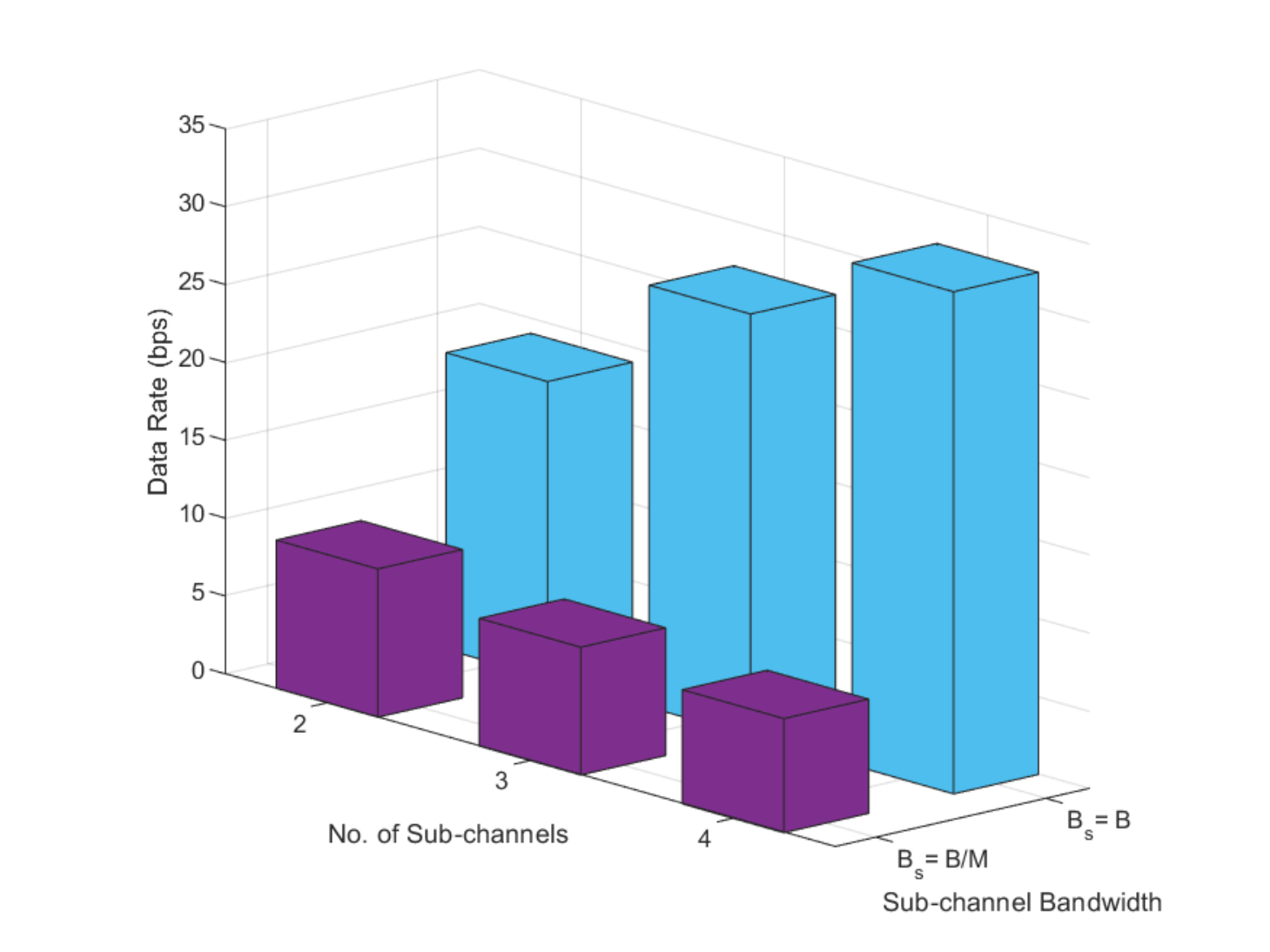}}
	\subfigure[No. of Sub-channels vs. No. of users]{\label{fig:7(b)}\includegraphics[scale=.6,keepaspectratio]{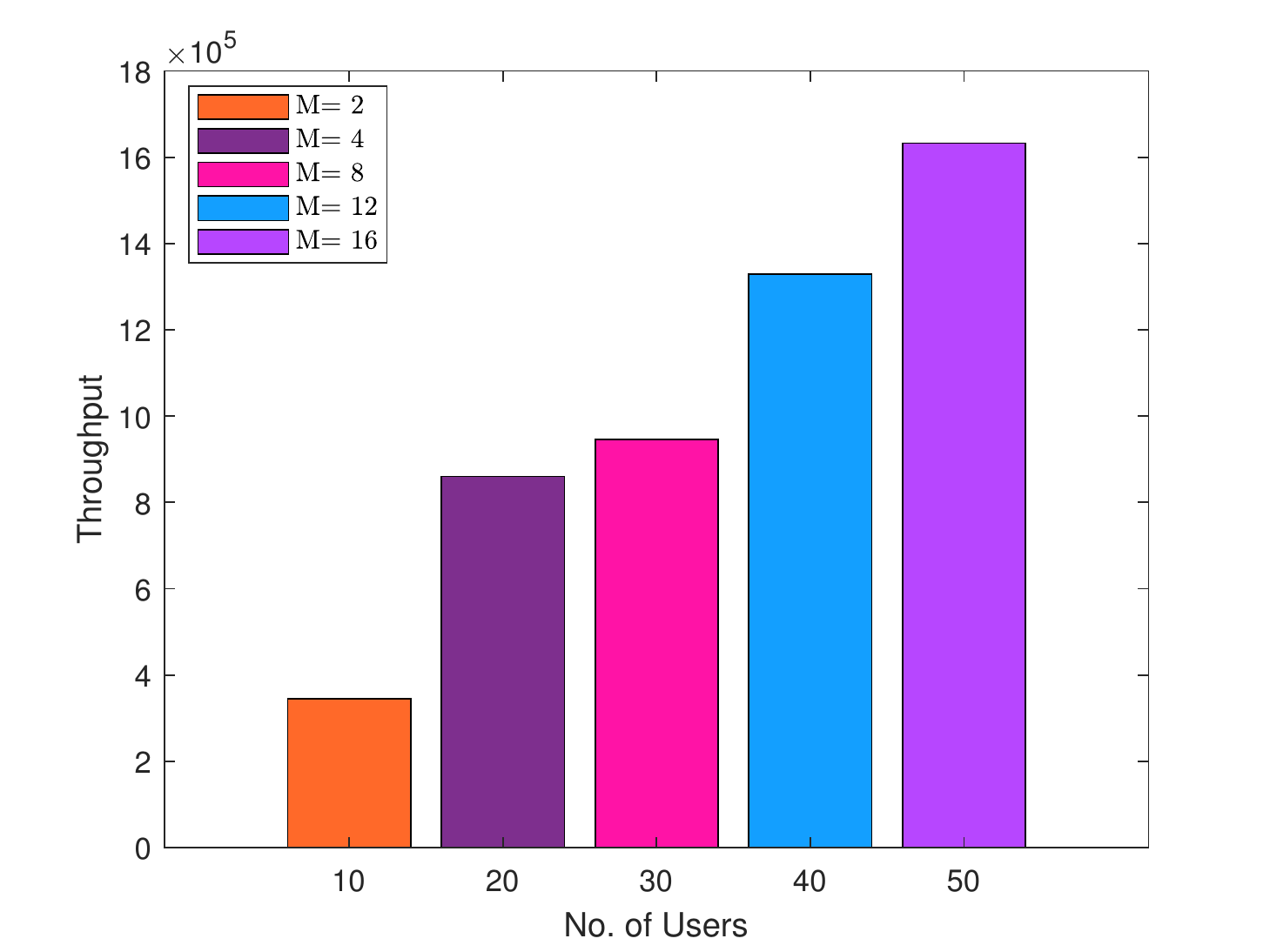}}
		\caption{{Spectral efficiency versus the number of sub-channels $M$: Sub-figure (a) shows two cases: First, The system bandwidth $B$ is equally divided into $M$ orthogonal sub-channels and thus each sub-channel has $B_s= \frac{B}{M}$ available bandwidth and secondly, bandwidth of each sub-channel $B_s = B$ . Sub-figure (b) represents system throughput w.r.t, increasing number of users and sub-channels.}}
	\label{subch}
	\vspace{-0.5 cm}
\end{figure}
\subsection{Impact of the System Density}
The number of sub-channels in a GF system affects the system performance by increasing or decreasing collision probability. Moreover, to achieve the Quality of Service (QoS) requirements for each IoT user and to maximize system throughput, the BS needs to find the optimal value of this parameter. The number of sub-channels is inversely proportional to the collision probability, but the increase in the number of sub-channel decreases the bandwidth of each sub-channel that results in poor spectral efficiency. On the other hand, increasing the number of sub-channels can enhance spectral efficiency subject to enough bandwidth. Fig. \ref{subch}\subref{fig:7(a)} shows the impact of the number of sub-channels on the system throughput.  Results are reported for two different scenarios, where different values of bandwidth is provided: $B_s= \frac{B}{M}$ and $B_s = B$. In the first case where the total bandwidth is equally divided into $M$ orthogonal sub-channels, it is clear that, with increasing number of sub-channels $M$, the system throughput decreases. It is worth noting that when the number of sub-channels is large, e.g., $M = 4$, the available bandwidth for each sub-channel is small and IoT users achieve a lower data rate over each sub-channel. Thus, IoT users in a network with a balanced number of sub-channels can achieve the threshold data rate ($R_{th}$) for successful SIC, and their QoS requirements can be satisfied. However, in the second case, when the system bandwidth increases, the spectral efficiency also increases. With limited bandwidth, system performance in terms of spectral efficiency is inversely proportional to the number of sub-channels. It is directly proportional to the number of sub-channels if the system bandwidth is high. Similarly, Fig. \ref{subch}\subref{fig:7(b)} shows the system throughput with increasing number of sub-channels (sub-channels with the same bandwidth i.e., 10 KHz) and GF users. A continuous increase in system throughput is reported as the number of users and sub-channels increases. Thus, with enough system bandwidth, system throughput can be increased with the increase in number of sub-channels.
\begin{figure}[t!]
	\centering
	\subfigure[Performance Comparison]{\label{set}\includegraphics[scale=.6,keepaspectratio]{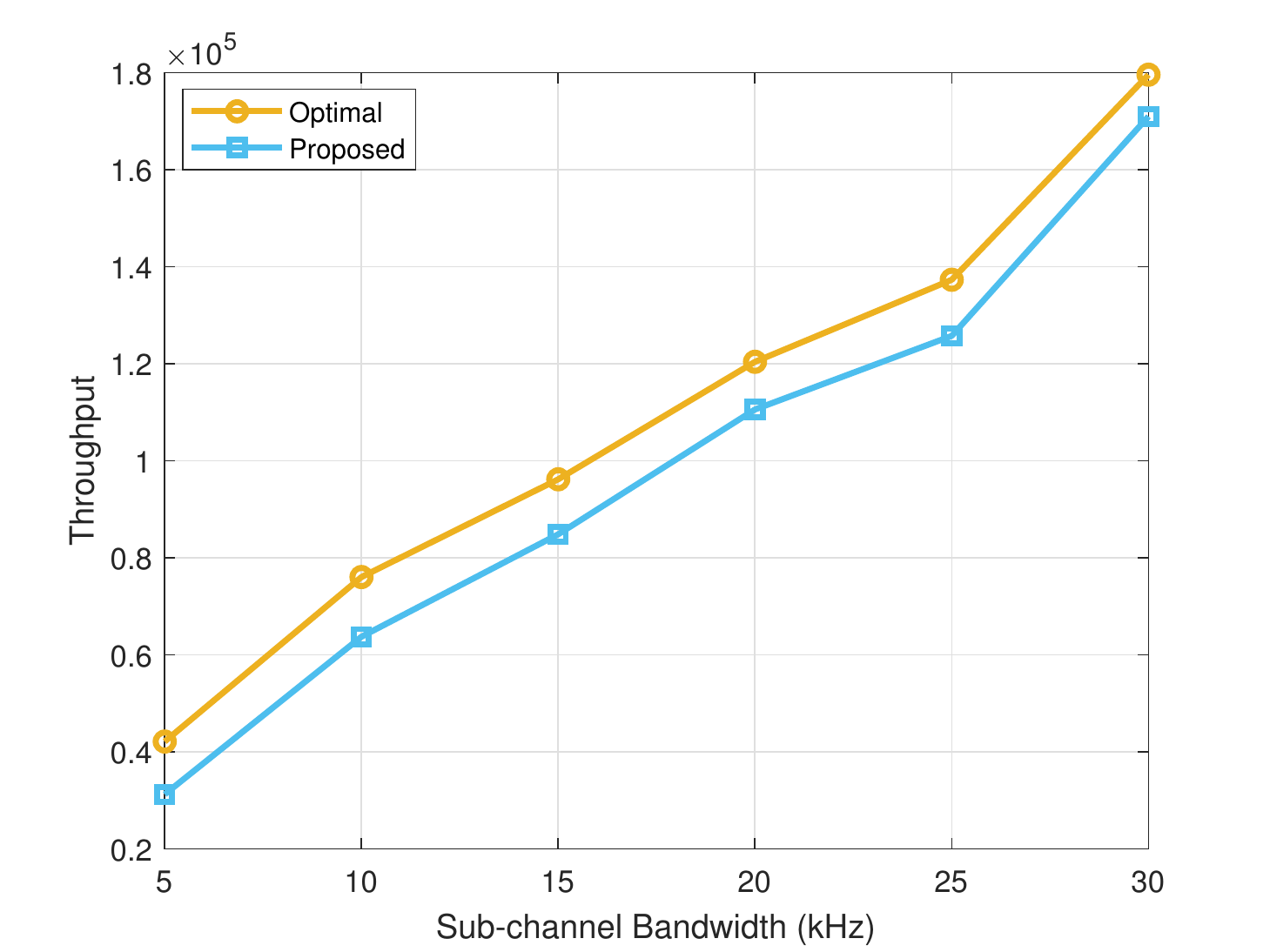}}
	\subfigure[Power allocation comparison]{\label{fix}\includegraphics[scale=.6,keepaspectratio]{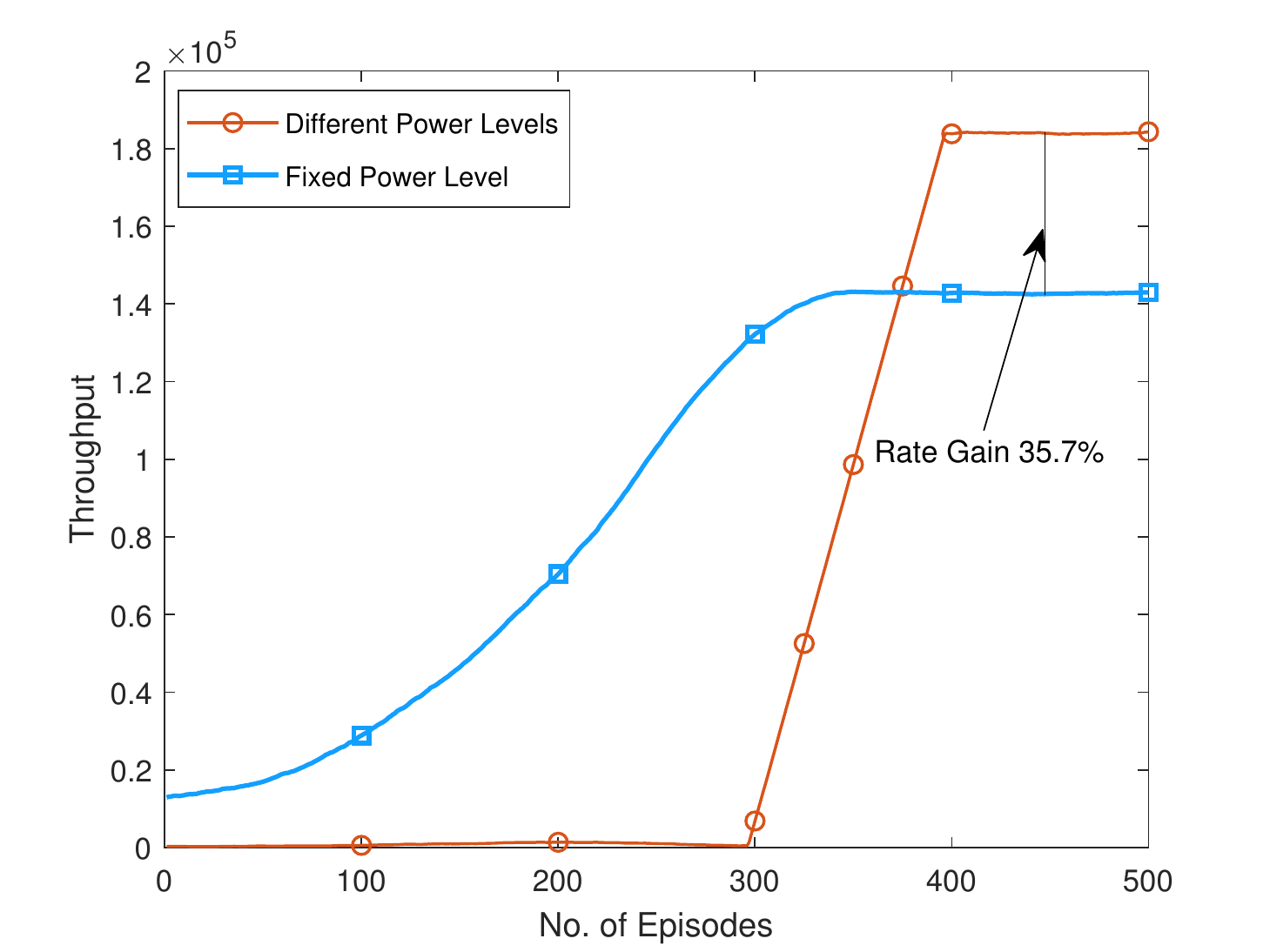}}
	\caption{{Performance and power allocation comparison with $M=4$ and $N_t=6$ (active users): Sub-figure (a) shows system performance with varying sub-channel bandwidth as compared to optimal solution. Sub-figure (b) represents the performance comparison in terms of throughput achieved by allocating fixed power to IoT users and power allocating from the available power pool.}}
	\label{fixed}
	\vspace{-0.5 cm}
\end{figure}

\subsection{Performance Comparison with Optimal solution}
When the number of variables in an optimization problem is small, the optimal solution can be found by exhaustive search. However, when there are many variables, finding an optimal solution to the resource allocation problem is extremely computationally expensive since it is known to be NP-hard. In this case, it is more reasonable to search for near-optimal solutions to reduce the computational complexity. The optimal resource allocation policy is based on exhaustive search that needs to explore/visit all network states, which is enormous in the case of IoT networks without learning policy. The Fig. \ref{fixed}\subref{set} shows the near optimal solution and the performance gap. In the proposed algorithm, the IoT user learns by interacting with the wireless environment and extracts the useful information for decision making (power and sub-channel selection). It is worth mentioning that visiting all the states of the network is not necessary for an agent (IoT user). The agent learns a strategy through the method of greedy exploration and exploitation. Therefore, we have achieved a near-optimal solution at the cost of linear time computational complexity.

\subsection{Fixed Power Allocation VS. Power Allocation from Available Power Pool}
A comparison of fixed power and power allocation from the power pool is shown in Fig. \ref{fixed}\subref{fix}. The network with the transmit power pool outperforms the network with fixed power allocation in terms of throughput. Algorithm with fixed transmit power level converges quickly due to the small number of action space. On the other hand, the algorithm with the transmit power pool requires more time to gain the highest throughput because of the large action dimensions. In fixed power allocation, users transmit with the highest power level resulting in increased interference to other users on the same sub-channel that leads to poor throughput gain, compromise user fairness and wastage of energy. Algorithm with available power pool achieves 35.7\% more throughput as compared to the fixed power allocation. This is because the algorithm dynamically allocates the transmit power to IoT users based on their communication distance and channel gain.
\begin{figure}[t!]
	\centering
	\subfigure[Multi-agent DQN based transmit power levels for each layer]{\label{DQNpower}\includegraphics[scale=.58,keepaspectratio]{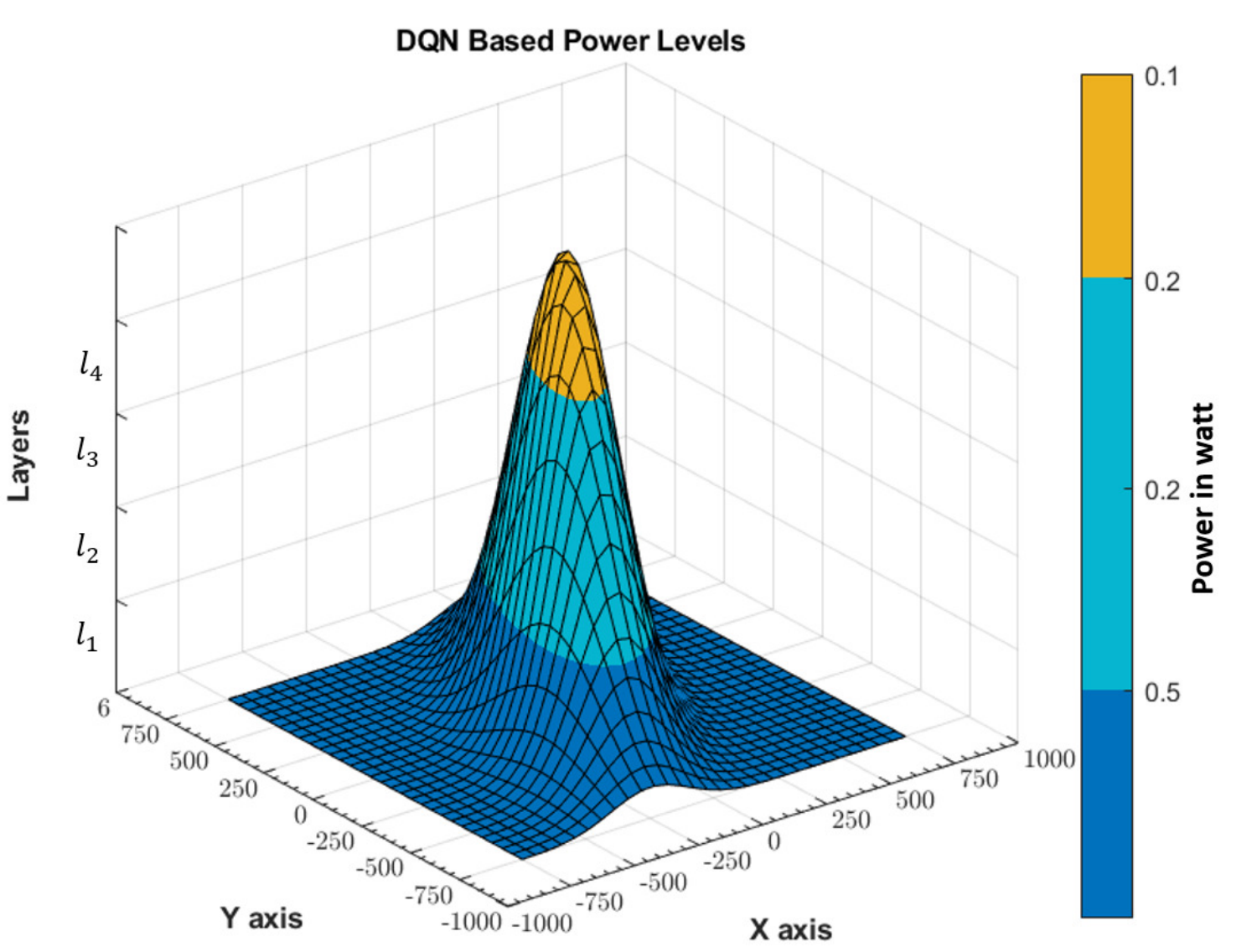}}
	\subfigure[Multi-agent DDQN based transmit power levels for each layer]{\label{DDQNpower}\includegraphics[scale=.58,keepaspectratio]{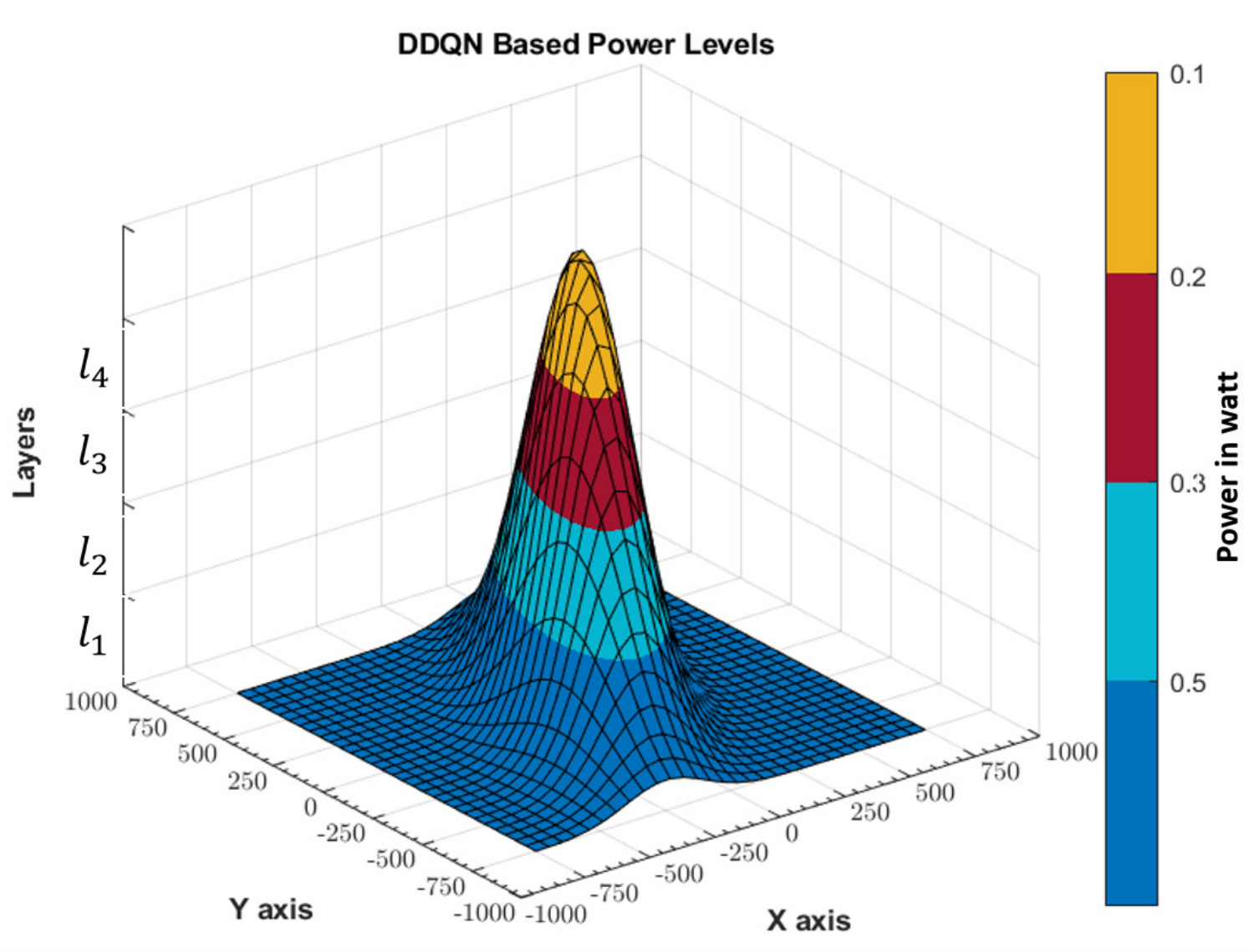}}
	\caption{Multi-agent DQN based transmit power levels vs. Multi-agent DDQN based transmit power levels for each layer: In Sub-figure (a) IoT users in layer 2 and 3 use same power level, whereas, in Sub-figure (b) IoT users in different layers select different transmit power levels for GF transmission that ensure sufficient received power difference at the BS.}
	\label{DQN_vs_DDQN}
	\vspace{-0.5 cm}
\end{figure}

\subsection{Multi-agent DRL based Prototype Power Pool}
In this section, we show the designed prototype power pool using multi-agent DRL algorithms under network settings $M=4$ and $N_t =8$ (active GF users).
\subsubsection{Multi-Agent DQN Based Prototype Power Pool}
Fig. \ref{DQN_vs_DDQN}\subref{DQNpower} shows the layer-based transmit power levels acquired via multi-agent DQN based GF-NOMA algorithm. It can be seen that users in layer $2$ and layer $3$ use the same transmit power levels.
It is evident from Fig.~\ref{cell_area}\subref{convergence} that DQN based GF-NOMA algorithm achieves less throughput as compared to DDQN. This is due to the overestimation of Q-values that converge our proposed algorithm with non-optimal actions (transmit power level selection). Moreover, in the power domain GF-NOMA, the received power level difference plays an important role in the decoding process. The transmit power of one user affects SINR and achievable capacity of other users on the same sub-channel, as users create interference to other users. In our proposed DQN based GF-NOMA algorithm, users in layer $2$ and $3$ use the same transmit power level that compromises NOMA fairness and achieve a suboptimal solution to the optimization problem mentioned in (\ref{of}).

\subsubsection{Multi-Agent Double DQN Based Prototype Power Pool}
To solve the issue of selecting the same transmit power level by IoT users in different layers, we proposed DDQN based GF-NOMA algorithm. In Fig. \ref{DQN_vs_DDQN}\subref{DDQNpower}, it can be observed that users select different power levels in different layers which is the motivation and objective of this research work. IoT users in layer 1 transmit with the highest power level, as the user with the highest received power at the BS face interference from users with weak received power levels in the decoding process. Moreover, to guarantee the SIC process, IoT users need to achieve the required data rate $R_{th}$, hence, users in layer 1 transmit data with the highest power level. Similarly, users in other layers obey the same process and BS decodes the users in the last layer (layer 4) without interference from other users. Therefore, users in the last layer can achieve the required data rate for successful SIC with a low transmit power level.
Furthermore, a performance difference in terms of energy consumption can be seen as users in layer-1 transmits with high power as compared to users in other layers. We will discuss a typical application scenario in terms of energy and time consumption with some specific protocols in our future work.
In addition, if the offline and online environment changes significantly, the proposed PP design may not suitable for online updating. Therefore, the proposed algorithm is suitable for networks with regular behaviours.

\subsection{Network Performance with and without Prototype Power Pool}
\begin{figure}[t!]
	\centering
	\subfigure[Convergence comparison]{\label{comp}\includegraphics[scale=.6,keepaspectratio]{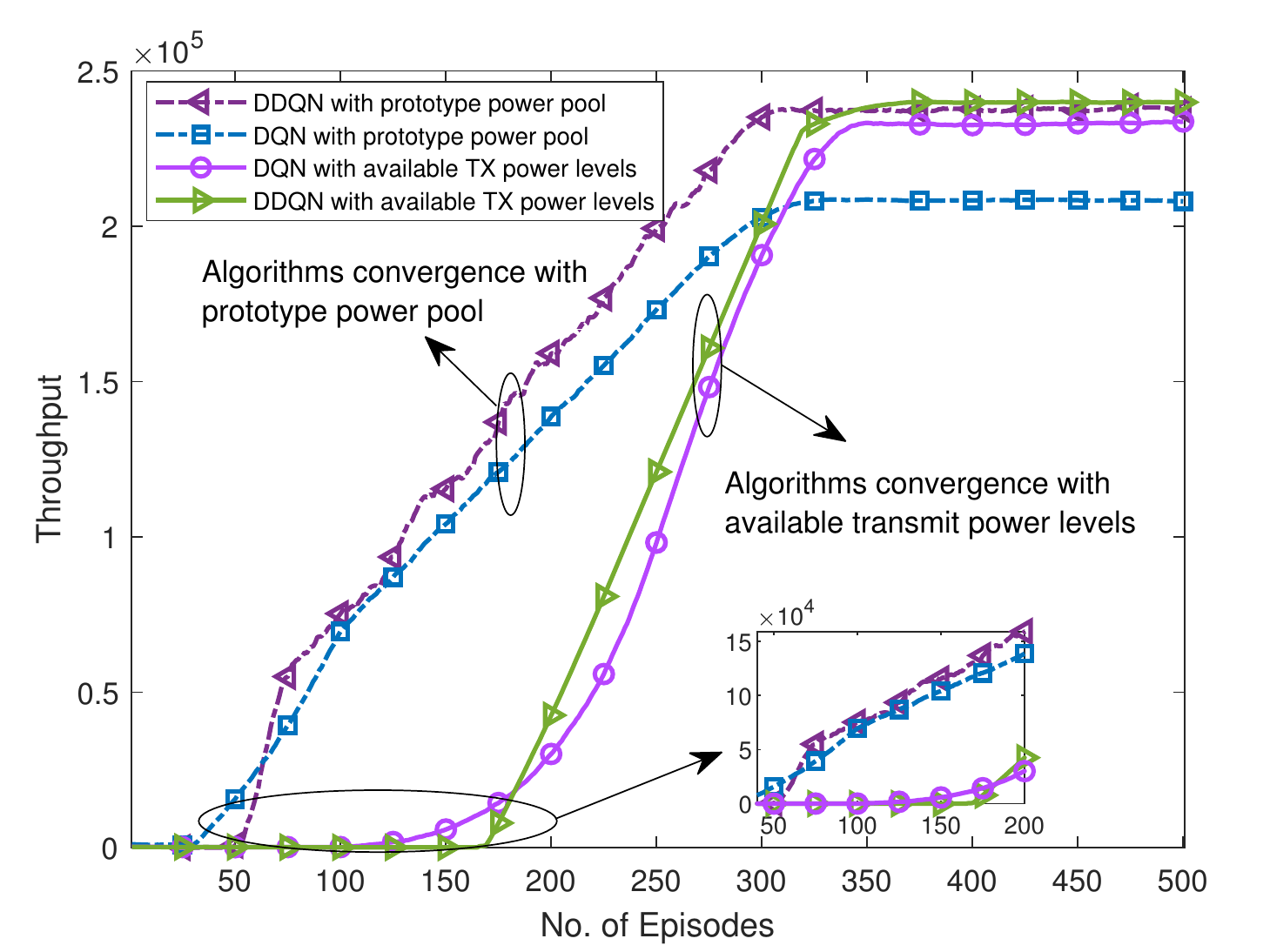}}
	\subfigure[Proposed GF-NOMA vs. GF-OMA]{\label{noma-oma}\includegraphics[scale=.6,keepaspectratio]{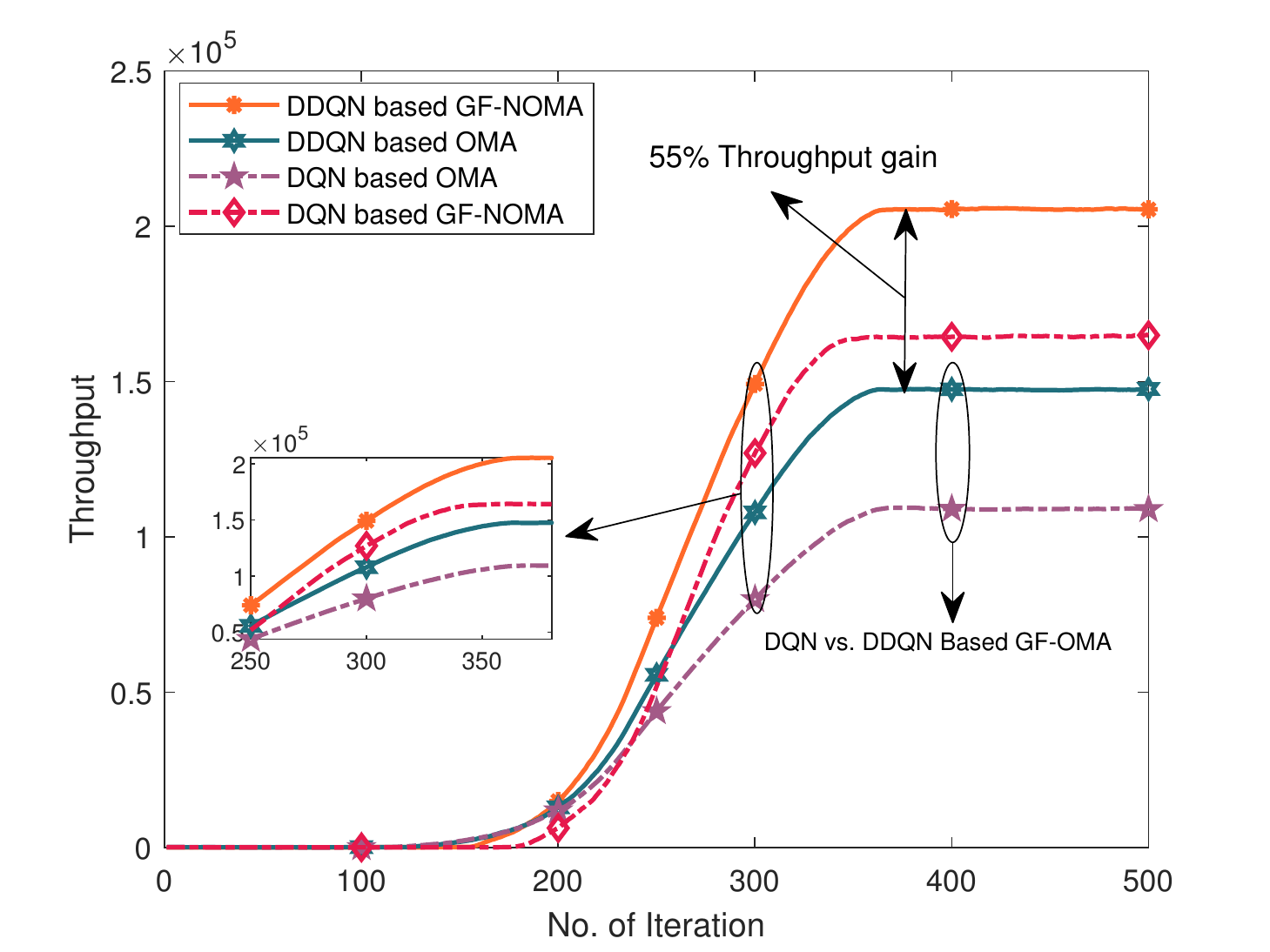}}
	\caption{Convergence comparison and performance of the proposed GF-NOMA vs. GF-OMA: Sub-figure (a) shows comparison of the proposed multi-agent DRL based GF-NOMA algorithms (multi-agent DQN based GF-NOMA and multi-agent DDQN based GF-NOMA) with and without the prototype power pool. Sub-figure (b) represents the comparison of proposed multi-agent DRL based GF-NOMA and GF-OMA. }
	\label{oma}
	\vspace{-0.5 cm}
\end{figure}

Fig. \ref{oma}\subref{comp} shows the convergence of the proposed multi-agent DDQN based GF-NOMA algorithm with the designed prototype power pool and available transmit power levels. It can be seen that the multi-agent DDQN and DQN based GF-NOMA algorithm with prototype power pool converges quickly with fewer training episodes. However, multi-agent DDQN algorithm outperforms multi-agent DQN algorithm in terms of throughput because DQN algorithm suffers from Q-values overestimation problem and converges with the non-optimal solution. On the other hand, agents in these algorithms with available transmit power levels require more training episodes to learn optimal actions due to an increase in action space.

\subsection{Proposed Multi-Agent DRL Based GF-NOMA VS. GF-OMA}
Fig. \ref{oma}\subref{noma-oma} shows the performance gain of the proposed multi-agent GF-NOMA IoT network over GF-OMA. It is evident that our proposed multi-agent DDQN based GF-NOMA algorithm outperforms GF-OMA with $55\%$ performance gain on system throughput. The reason behind this performance gain is the splitting of bandwidth resources among the IoT users in OMA. Since users (agents) in multi-agent GF-NOMA select power level and sub-channel (actions) in such a way that maximizes the throughput (reward). Therefore, an accurate power allocation and grouping of users with significant channel gain difference in a NOMA cluster result in performance gain over GF-OMA. Furthermore, the throughput achieved by multi-agent DDQN based GF-NOMA and multi-agent DDQN based GF-OMA is superior to multi-agent DQN based GF-NOMA and multi-agent DQN based GF-OMA.
\section{Conclusion}
This paper has generated a layer-based transmit power pool for GF-NOMA IoT networks by utilizing multi-agent DRL.  We have divided the cell area into different layers and calculated the optimal transmit power level for each layer to ensure sufficient received power difference at the BS to maximize connectivity. Moreover, numerical results have shown that the multi-agent GF-NOMA algorithm outperforms conventional GF-OMA in terms of throughput. The prototype power pool design has been proved to perform better than the fixed power allocation design and pure online training. Finally, we have identified the transmit power levels for the prototype power pool.
Investigating the proposed algorithm for multi-antenna scenarios and considering user fairness (in terms of energy consumptions) are some of the promising future directions. Additionally, to investigate environmental impact on the considered DRL algorithm is another possible extension to this work.

\vspace{-0.5cm}
\bibliographystyle{IEEEtran}
\bibliography{mybib}
\end{document}